\documentclass{ieeeaccess}
\usepackage{colortbl}
\usepackage{tikz}
\usepackage{pgfplots}
\pgfplotsset{compat=1.18}
\usepackage{tcolorbox}
\usepackage{amsmath,amssymb,amsfonts}
\usepackage{mathtools}
\usepackage{algorithmic}
\usepackage{algorithm}
\usepackage{array}
\usepackage[caption=false,font=normalsize,labelfont=sf,textfont=sf]{subfig}
\usepackage{textcomp}
\usepackage{stfloats}
\usepackage{hyperref}
\usepackage{xurl}
\hypersetup{breaklinks=true}
\usepackage{verbatim}
\usepackage{booktabs}
\usepackage{graphicx}
\usepackage{multirow}
\usepackage{tabularx}
  \newcolumntype{Y}{>{\centering\arraybackslash}X}
\usepackage{makecell}
\usepackage{standalone}
\usepackage{enumitem}
\usepackage{listings}
\usepackage{microtype}
\usepackage[numbers,sort&compress]{natbib}
\usepackage{multicol}
\usepackage{xcolor}
\usetikzlibrary{positioning, arrows, arrows.meta, calc, automata, shapes, shadows, fit, backgrounds}
\usepackage[T1]{fontenc}
\usepackage[utf8]{inputenc}
\usepackage[nolist,nohyperlinks]{acronym}
\usepackage{xspace}

\DeclareFontShape{T1}{formata}{m}{sl} { <-> ssub * formata/m/it }{}
\NewSpotColorSpace{PANTONE}
\AddSpotColor{PANTONE} {PANTONE3015C} {PANTONE\SpotSpace 3015\SpotSpace C} {1 0.3 0 0.2}
\SetPageColorSpace{PANTONE}

\makeatletter
\AtBeginDocument{\DeclareMathVersion{bold}
\SetSymbolFont{operators}{bold}{T1}{times}{b}{n}
\SetSymbolFont{NewLetters}{bold}{T1}{times}{b}{it}
\SetMathAlphabet{\mathrm}{bold}{T1}{times}{b}{n}
\SetMathAlphabet{\mathit}{bold}{T1}{times}{b}{it}
\SetMathAlphabet{\mathbf}{bold}{T1}{times}{b}{n}
\SetMathAlphabet{\mathtt}{bold}{OT1}{pcr}{b}{n}
\SetSymbolFont{symbols}{bold}{OMS}{cmsy}{b}{n}
\renewcommand\boldmath{\@nomath\boldmath\mathversion{bold}}}
\makeatother

\definecolor{blue}{RGB}{5, 107, 138}
\colorlet{blue}{blue!80}
\definecolor{red}{RGB}{187, 68, 48}
\definecolor{green}{RGB}{246, 174, 45}

\newcommand{\imagedir}{./images/}

\newcommand{\nmax}{\ensuremath{n_\text{max}}\xspace}
\newcommand{\nmin}{\ensuremath{n_\text{min}}\xspace}

\newif\ifanonymous
\anonymousfalse

\renewcommand{\thevol}{X}
\renewcommand{\theyear}{X}

\begin{document}

\doi{}

\title{Embedding Large PQC KEMs in IKEv2}

\ifanonymous
  \author{Anonymous Author}
  \address{Anonymous Institution}
\else
\author{
  \uppercase{Daniel Herzinger}\authorrefmark{1},
	\uppercase{Marco Thomas}\authorrefmark{1},
	\uppercase{Tobias Handirk}\authorrefmark{1},
  \uppercase{Stefan-Lukas Gazdag}\authorrefmark{1},
  \uppercase{Daniel Loebenberger}\authorrefmark{2},
	\uppercase{Marian Margraf}\authorrefmark{2}
}

\address[1]{genua GmbH, Kirchheim, 85551 Germany}
\address[2]{Fraunhofer AISEC, Garching, 85748 Germany}

\corresp{Corresponding author: Daniel Herzinger (e-mail: daniel\_herzinger@genua.de)}

\tfootnote{Part of this work was funded by the German Federal Ministry of Education and Research respectively the German Federal Ministry of Research, Technology and Space in the projects AMiQuaSy (project ID 16KIS1817) and SUSTAINET (project ID 16KIS2242 under the CELTIC-next program).}
\fi

\begin{abstract}
Post-quantum key encapsulation mechanisms carry public keys far larger than the established VPN handshake protocol IKEv2 was designed for.
At the same time, its initial exchange comprising the public keys cannot be fragmented. In response, the IETF discusses a range
of protocol adjustments, yet these proposals have never been compared in terms of their performance
and stability under identical conditions. We systematically screened the IETF
document corpus for such proposals, added three of our own (one using forward error correction
and two using IKEv2 with QUIC, namely QUIKE and QUIKE with a composite key exchange), identified
two specification gaps in existing acknowledgment drafts, and implemented all eight on the same
open-source codebase. Measurements across different bandwidths, latencies, and packet loss rates,
and with ML-KEM, FrodoKEM, and Classic McEliece, total $126\,500$ handshakes in $1296$ different
configurations and show that no mechanism fully compensates for the effect of the key size on
performance. For Classic McEliece, no approach handles challenging conditions acceptably. The
approaches either congest the link or take up to nearly two minutes per handshake on average.
For ML-KEM and FrodoKEM, there are two ways forward: building on already standardized
extensions or delegating the transport. The former requires the larger
implementation change and needs further optimization, such as acknowledgment mechanisms or 
error-correcting codes, but grants continuity and compatibility. The latter can use a composite key
exchange to avoid these drawbacks and achieve a faster, more resilient handshake and leaner 
implementation, with QUIKE-composite offering the best overall performance.
\end{abstract}
\vspace{.5cm}

\begin{keywords}
  Cryptography, Forward error-correcting codes, IKEv2, IPsec, PQC, QUIC, VPN
\end{keywords}

\titlepgskip=-21pt
\maketitle

\newcounter{savefootnote}
\setcounter{savefootnote}{\value{footnote}}

\renewcommand{\thefootnote}{\relax}%
\footnotetext{This work has been submitted to the IEEE for possible publication. Copyright may be transferred without notice, after which this version may no longer be accessible.}%

\setcounter{footnote}{\value{savefootnote}}
\renewcommand{\thefootnote}{\arabic{footnote}}

\begin{acronym}
  \acro{ADDKE}{Additional Key Exchanges}
  \acro{AEAD}{Authenticated Encryption with Associated Data}
	\acro{DH}{Diffie-Hellman}
	\acro{DoS}{Denial of Service}
	\acro{EAP}{Extensible Authentication Protocol}
	\acro{FEC}{Forward Error-correcting Codes}
	\acro{ICV}{Integrity Check Value}
	\acro{KDF}{Key Derivation Function}
	\acro{KEM}{Key Encapsulation Mechanism}
	\acro{LDPC}{Low Density Parity Check}
	\acro{LT}{Luby Transform}
	\acro{MTU}{Maximum Transmission Unit}
	\acro{MSS}{Maximum Segment Size}
	\acro{NIST}{National Institute of Standards and Technology}
	\acro{PMTU}{Path \ac{MTU}}
	\acro{PQC}{Post-Quantum Cryptography}
	\acro{PSK}{Pre-Shared Key}
	\acro{PoC}{Proof of Concept}
	\acro{QC}{Quantum Computer}
	\acro{QKD}{Quantum Key Distribution}
	\acro{Raptor}{Rapid Tornado}
	\acro{RTT}{Round-Trip Time}
	\acro{SA}{Security Association}
	\acro{VPN}{Virtual Private Network}
\end{acronym}

\section{Introduction}%
\label{sec:introduction}
\acp{VPN} securely interconnect IT infrastructure over public networks like the
Internet, representing a backbone for modern distributed systems. However, the
prospect of a large \ac{QC} threatens the asymmetric cryptography building blocks
vital for establishing secure \ac{VPN} connections. Even more urgently, an attacker
can perform \emph{store-now, decrypt-later} attacks, recording \ac{VPN} connections
today and using a \ac{QC} to break the key agreement and decrypt the data in the
future.

New \ac{PQC} key agreement mechanisms also protect against the \ac{QC} threat.
They typically consist of \acp{KEM}~\cite{holding_kems}. Compared to classical
\ac{DH} constructs, they use a new syntax and both their public keys and
ciphertexts are drastically larger. This complicates integrating them into the
prevalent \ac{VPN}
protocols designed and optimized for classical cryptography. One example is
IPsec's handshaking protocol IKEv2~\cite{ikev2}, first published in 2005. Its
IETF working group acknowledged the need for adaptation, publishing the
first \ac{PQC} key agreement RFC~\cite{rfc9370} in 2023 and discussing further
optimizations and alternatives since. Yet, these approaches are complex and face
reliability issues, especially considering large \ac{PQC} \acp{KEM} and difficult
network environments~\cite{real_ipsec, selective_retransmission}.

Such a fundamental change to a \ac{VPN} protocol needs to balance requirements in
terms of, but not limited to, performance, resilience, backwards compatibility,
complexity, and implementation intrusiveness. The resulting situation is complex,
exacerbated by a lack of comprehensive data even regarding the well-quantifiable
dimensions of performance and network resilience. In response, we conducted a systematic
search for all relevant proposals for \ac{PQC} \ac{KEM} protocol adjustments for IKEv2
discussed at the IETF\@.
Additionally, we propose three novel approaches: one using \ac{FEC} and two
versions of \emph{QUIKE}, where we combine IKEv2 and QUIC\@. We evaluate all
proposals in terms of performance and network resilience in one stable test
environment. This yields clear and reproducible data for a fair comparison and
discussion of the qualitative advantages and drawbacks of both the drafts already in
discussion as well as our novel
ones. It shows that for the largest \acp{KEM}, no approach offers acceptably stable
performance. For smaller and medium-sized \acp{KEM}, two pathways arise: One
follows the already standardized way, adhering to design decisions and keeping
continuity, but requires more complex adjustments and optimizations to provide
performance and stability. The other moves away from existing design decisions
and breaks compatibility, but offers the highest and most stable performance with an
overall simpler design, especially QUIKE with a composite key exchange. The results
act as guidance and clarification for the standardization working group, the academic
community, and implementers making trade-offs within their software. We enable this
with the following contributions:
\begin{itemize}
	\item A systematic literature review across the IETF document corpus, identifying
		proposals for \ac{PQC} \acp{KEM} in IKEv2.
	\item An approach to stabilize the transmission of fragmented large \ac{KEM} keys
		in IKEv2 using \ac{FEC}.
	\item Two approaches which fuse IKEv2 and QUIC~\cite{quic} into QUIKE, a fast and
		reliable VPN key exchange protocol for IPsec; one using the previous
		``one-ke-per-round'' semantics and one using a composite key agreement.
	\item The identification of two specification gaps, concerning key retention and
		retransmission limits, in drafts adding acknowledgments to IKEv2 fragmentation,
		and a defect in the QUIC library we build on.
	\item A reproducible evaluation of the handshake time of all approaches under varying
		latency, bandwidth, and packet loss as well as with three different \acp{KEM}.
	\item A qualitative discussion of the approaches' overall complexity,
		implementation intrusiveness, and practicability.
\end{itemize}

The remainder of this paper is structured as follows:
Section~\ref{sec:background} gives an overview of IKEv2, \ac{PQC} \acp{KEM}, \ac{FEC}, QUIC, and related
work.
Section~\ref{sec:methodology} states our research questions and describes both the
systematic search and the measurement setup. Section~\ref{sec:ikev2_pqc} presents the
screened proposals and our three novel approaches; Section~\ref{sec:evaluation}
covers their implementation and the measurement results, which
Section~\ref{sec:discussion} puts into perspective. Finally,
Section~\ref{sec:threats_to_validity} explores possible threats to the validity of
our study and Section~\ref{sec:conclusion} concludes.

\section{Background and Related Work}%
\label{sec:background}
This work focuses on \ac{PQC} \ac{KEM} integration into IKEv2, the current
handshaking protocol of the IPsec suite. The other protocol widely used in IPsec is
ESP~\cite{esp_rfc}. It handles the payload encryption in the connections established
by IKEv2. It relies on symmetric encryption schemes which, albeit not unaffected by
\acp{QC}, are still able to provide sufficient security properties for those
use cases. Thus, this paper concentrates on IKEv2 which is severely affected by the
\ac{QC} threat.

\subsection{Classic IKEv2}%
\label{sec:bg_ikev2}
IKEv2 is a UDP-based handshaking protocol defined in RFC~7296~\cite{ikev2}. It is
part of the IPsec protocol suite and negotiates the used cryptographic parameters,
derives the keys for symmetric encryption, and authenticates the connection. After
completing a handshake, it installs \emph{\acp{SA}}: data structures containing all
necessary information, including the keys, for handling the encrypted traffic of one
connection. They are divided into bidirectional \emph{IKE \acp{SA}} which hold the
information for the IKEv2 connection itself, and unidirectional \emph{Child \acp{SA}}
with the information for the actual payload data to be transmitted inside the
\ac{VPN} via ESP\@. Hence, after the IKEv2 handshake finishes, there is one IKE
\ac{SA} and a pair of Child \acp{SA} per peer.

\begin{figure}[tb]
	\centering
	\providecommand{\imagedir}{./}
\input{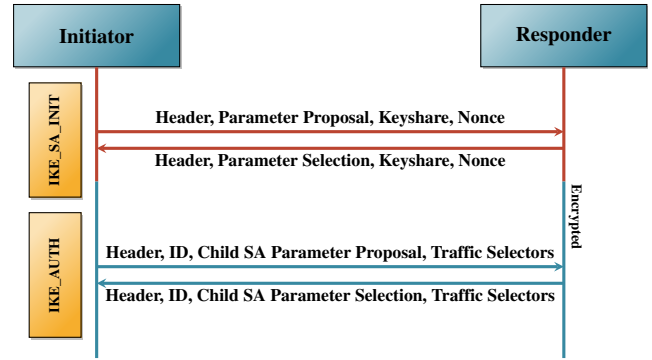}

\begin{tikzpicture}
	\def\width{450 pt}
	\def\rowsep{130}		
	\def\plainrun{110}		
	\def\encrun{170}		

	\node (init) [blue_box] {\huge\textbf{Initiator}};
	\node (resp) [blue_box] at ([xshift=\width] init)
		{\huge\textbf{Responder}};

	\draw[lifeline, draw=red] (init.south) -> ++(0,-\plainrun pt) coordinate(i1);
	\draw[lifeline] (i1) -> ++(0,-\encrun pt);
	\draw[lifeline, draw=red] (resp.south) -> ++(0,-\plainrun pt) coordinate(r1);
	\draw[lifeline] (r1) -> ++(0,-\encrun pt) node[ sloped, pos=0.2,
		above] {\bfseries\Large Encrypted};

	\coordinate (sa_init) at ([yshift=-70]init.south);
	\node[exchange] at ([xshift=-40]sa_init) {\bfseries\Large IKE\_SA\_INIT};

	\draw[arrow, draw=red] ([yshift=8]sa_init) -> ++(\width,0)
		node[pos=0.5, above] {\bfseries\LARGE Header, Parameter Proposal,
		Keyshare, Nonce};
	\draw[arrow, stealth-, draw=red] ([yshift=-8]sa_init) -- ++(\width,0)
		node[pos=0.5, below] {\bfseries\LARGE Header, Parameter Selection,
		Keyshare, Nonce};

	\coordinate (ike_auth) at ([yshift=-\rowsep]sa_init);
	\node[exchange, minimum width=120] at ([xshift=-40]ike_auth)
		{\bfseries\Large IKE\_AUTH};

	\draw[arrow] ([yshift=8]ike_auth) -> ++(\width,0)
		node[pos=0.5, above] {\bfseries\LARGE Header, ID, Child SA
		Parameter Proposal, Traffic Selectors};
	\draw[arrow, stealth-] ([yshift=-8]ike_auth) --
		++(\width,0) node[pos=0.5, below] {\bfseries\LARGE Header, ID,
		Child SA Parameter Selection, Traffic Selectors};

\end{tikzpicture}%
	\caption{The message flow of a classic IKEv2 handshake.}%
	\label{fig:ikev2}
\end{figure}

IKEv2 itself is structured into \emph{exchanges}. In each exchange, both
parties, the \emph{initiator} and the \emph{responder}, each send one message.
Figure~\ref{fig:ikev2} illustrates the handshake structure. The first exchange
(called \emph{IKE\_SA\_INIT}) starts the handshake, carries the negotiation of
parameters and compatibility with possible IKEv2 extensions, and exchanges
nonces and a first shared secret. All subsequent exchanges are then encrypted.
The next exchange, i.e., \emph{IKE\_AUTH}, handles identity information,
proposals for the first Child \acp{SA}, and a signature authenticating the
handshake. Most subsequent actions, e.g., rekeying or creating additional Child
\acp{SA}, are done via the \emph{CREATE\_CHILD\_SA} exchange.

To be as lean as possible and have no unnecessary round-trips, IKEv2 uses UDP as its
base. If packets get dropped on their path, the initiator notices that there is no
answer from the responder which triggers a retransmission of their last exchange
message. The responder, upon receiving a retransmitted message, assumes their answer
got lost and retransmits theirs. As in classical setups most messages are small and
the handshake is over after usually only two exchanges, transport features like
congestion control are not considered necessary.

One real limitation that comes with UDP is that no message can exceed the
\ac{MTU} of the two \ac{VPN} devices and of any of the middleboxes on the packet's
network path, which typically is at most $1500$ bytes, sometimes $1280$ bytes or
less. Theoretically, IP fragmentation should solve this issue. However, that
is unreliable in practice and, thus, must be avoided~\cite{ike_frag}.
RFC~1191~\cite{pmtu} gives techniques to determine the \ac{PMTU}, yet as soon as an
exchange message exceeds it, the handshake fails. To still be able to send, e.g.,
large certificate chains in IKE\_AUTH, RFC~7383~\cite{ike_frag} defines fragmentation
on the IKEv2 level. However, it must not be used for the IKE\_SA\_INIT
exchange~\cite{ike_frag} which is a major challenge for incorporating large \ac{PQC}
\acp{KEM}.

\subsection{Post-Quantum KEMs}%
\label{sec:bg_kems}
Classical key agreement usually uses \ac{DH} structures. There, both parties
agree on public prime-order group $\mathbb{G}$ with generator $g$ and compute their key shares as $g^x$
and $g^y$ for their uniformly random secrets $x$ and $y$, respectively. Upon receiving
the other party's key share, they compute $(g^y)^x$ and $(g^x)^y$,
resulting in the same shared secret on both sides. However, if an attacker is able to solve the discrete logarithm
problem in $\mathbb{G}$, it is able to compute the shared secret $g^{xy}$ from the public key shares 
$g^x$ and $g^y$. While there is no known efficient algorithm for calculating a
solution for it on classical computers, a large \ac{QC} is expected to be able to
solve it in polynomial time~\cite{shor,bernstein2017post}.

In response to the \ac{QC} threat, the \ac{NIST} started a competition to find the
best quantum-safe, hence \ac{PQC}, key agreement replacement focusing on KEMs~\cite{nist_pqc_cfp}.
This means that one party generates a
public/private key pair $KeyGen() \rightarrow p, s$ and sends their ephemeral public
key $p$ to their peer. The peer encapsulates a random secret key $k$ using the
initiator's public key $Encaps(p) \rightarrow k, c$ and sends the ciphertext $c$ back to the
initiator who uses their secret key $s$ to also obtain the shared secret $Decaps(c,
s) \rightarrow k$. Apart from the new syntax, \acp{KEM} also have new properties that
determine their security against chosen plain- or ciphertext attacks which need to be
taken into account to prevent introducing new vulnerabilities~\cite{holding_kems}.

\begin{table}[!tb]
	\caption{\acs{NIST} security levels~\cite{nist_pqc_cfp}.}%
\label{tab:nist_levels}
\centering
\begin{tabularx}{\columnwidth}{lX}
\toprule
\textbf{Level} & \textbf{Comparable difficulty} \\
\midrule
1 & Comparable to a key search on AES-128. \\
2 & Comparable to a collision search on SHA3-256. \\
3 & Comparable to a key search on AES-192. \\
4 & Comparable to a collision search on SHA3-384. \\
5 & Comparable to a key search on AES-256. \\
\bottomrule
\end{tabularx}
\end{table}

The \acp{KEM} rely on the hardness of different mathematical problems in the field of
unstructured or semistructured lattices and linear codes with errors, which are all
assumed to be hard to solve, even with a \ac{QC}. Yet, as cryptanalysis is not as
advanced regarding them, they are still often only used in hybrid mode, i.e., in combination
with a classical key agreement, to prevent the risk of actually downgrading
security~\cite{bsi_crypto,bsi_handlung}. Combining the classical and PQC shared secrets into one key is a well-studied
problem with ready-made constructions~\cite{x_wing,holding_kems}. To classify the different algorithms and their parameters and
give guidance regarding their security, \ac{NIST} put them into categories, shown in
Table~\ref{tab:nist_levels}.

\begin{table}[!tb]
	\caption{Sizes of the transmitted objects, in bytes.}%
\label{tab:kem_sizes}
\centering
\begin{tabular}{llrr}
\toprule
\textbf{Scheme} & \textbf{Level} & \textbf{Public key} &
	\textbf{Ciphertext} \\
\midrule
\multicolumn{4}{l}{\textit{Classical key agreements}} \\
\ac{DH} MODP-3072     & 0 &                384 & 384\textsuperscript{*}    \\
\ac{DH} MODP-8192     & 0 &							1\,024 & 1\,024\textsuperscript{*} \\
X25519								& 0 &                 32 & 32\textsuperscript{*}     \\
\midrule
\multicolumn{4}{l}{\textit{Post-quantum KEMs}} \\
ML-KEM-768               & 3 &            1\,184 &  1\,088 \\
ML-KEM-1024              & 5 &            1\,568 &  1\,568 \\
FrodoKEM-976						& 3 &           15\,632 & 15\,744 \\
FrodoKEM-1344						& 5 &           21\,520 & 21\,632 \\
Classic McEliece 460896  & 3 &  524\,160 &      156 \\
Classic McEliece 8192128 & 5 & 1\,357\,824 &      208 \\
\bottomrule
\end{tabular}

\vspace{2pt}
\footnotesize
\textsuperscript{*}~\ac{DH} constructs send key shares with the same structure in
both directions.
\end{table}

The different syntax and properties are mostly an implementation task. The
biggest protocol-related difference is the \ac{PQC} \acp{KEM}' sizes.
Table~\ref{tab:kem_sizes} compares prevalent classical \ac{DH} key agreements with
different \ac{PQC} \acp{KEM}. It shows the symmetrical nature of \ac{DH}
constructs, while the \acp{KEM} sometimes are highly asymmetrical with the initiator
usually transferring more data than the responder. The other big aspect for protocol
design is the total size. A protocol which requires its first message to fit the
\ac{MTU}, i.e., $1500$ bytes by default, sometimes less, needs modification to
support almost any \ac{PQC} \ac{KEM}. We further show that the rudimentary
fragmentation mechanism of RFC~7383~\cite{ike_frag} is insufficient for dealing with
packet loss or congestion at large \ac{KEM} sizes.
A Classic McEliece public key exceeds the size of an IKEv2 header length field,
requiring even further modifications.

Cryptanalysis advances published in concurrent
work~\cite{mceliece_keyrecovery,cryptoeprint:mceliece-dist} reduce the
long-term trust in Classic McEliece's security~\cite{bsi_mceliece}. Still, it
represents a good example for evaluation in the upper \ac{KEM}-size regime, as
similar advances might also affect other \acp{KEM}, potentially driving up the
required \ac{KEM} sizes across the schemes. Deployments with high security
requirements and \emph{store-now, decrypt-later} attacks in mind must take the risk
of further (quantum-)cryptanalysis advances into account and require larger, more
conservative \acp{KEM}.

\subsection{Forward Error-Correcting Codes}%
\label{sec:bg_fec}
\ac{FEC} is a well-established concept in the world of data transmission over lossy channels and storage media. \ac{FEC} encodes source blocks and emits
redundant encoded blocks, so the receiver does not need all of them to recover the data. The
decoding success depends on the coding and loss pattern of the channel. In a \ac{VPN} context, this
recovery allows message reconstruction despite packet loss, minimizing handshake failure. The codes manifest in three primary groups: linear block codes, convolutional codes, and
fountain codes.

\begin{description}
\item[Linear block codes.]
Linear block codes generally possess the characteristics of being of a fixed rate. This means
that once encoding parameters are established for a chunk of data, the resulting code is
deterministically sized and cannot change in length or recovery power.

\emph{Reed-Solomon codes}~\cite{Reed1960} from 1960 are well-known due to
their simplicity and being a maximum distance separable (MDS)
code~\cite{Roman1992Coding,MacWilliams1977}. They use a fixed block sharding
approach, where the quantity of excess shards determines recoverability. The shard
size $s$ is arbitrary but uniform across all shards. After chunking the source data
into $k$ shards of size $s$, a coding matrix yields the code. If at least $k$
shards arrive at the destination, the message has a decoding probability of $100\%$
and $0\%$ otherwise.

Also the probabilistic \emph{\ac{LDPC}} code by Gallager from
1962~\cite{gallager1962,borwankarLow2020,nasaLdpc} is well-established in 5G radio
access technology~\cite{robert2016}.

\item[Convolutional codes.]
Convolutional codes do not encode fixed-length blocks independently, but rather
generate each output symbol from a sliding window of the input stream. \emph{Turbo
codes}, introduced by Berrou et al.\ in 1993~\cite{berrou1993,Benedetto1995}, are
built from such codes: they are parallel concatenations of two recursive
convolutional encoders. In practice, they are applied to fixed-size frames, as the
interleaver and the iterative decoder operate on a whole block. Turbo codes achieve
an error-correction performance close to the Shannon limit~\cite{shannon1948}, but at
the cost of iterative decoding. Further, they target noisy channels rather than pure
erasure channels: the two encoders are separated by an interleaver, which
decorrelates their inputs so that low-weight codewords are unlikely to occur in both
encoders and the decoders can exchange largely independent information. This also
spreads burst errors over the codeword. Rowshan et al.~\cite{rowshan2024} show for the
upcoming 6G mobile standard that the interleaver design has a substantial impact on
the correction performance, and outline different designs for such codes.

\item[Fountain codes.]
Fountain codes, also called rateless codes, are the newest class of \ac{FEC}.
Luby et al.~\cite{luby1997} published a precursor, the \emph{tornado codes}, in 1997.
They work by layering \ac{LDPC} and Reed-Solomon codes to yield the best of both
worlds in terms of efficiency and recovery chance. The result requires more symbols
than Reed-Solomon, but performs many orders of magnitude faster~\cite{byers1998}.
After refinement, Luby published \emph{\ac{LT}} codes~\cite{luby2002} in 2002. This
represents the first version of modern fountain codes.

Both \emph{\ac{Raptor}}~\cite{rfc5053} and its successor \emph{RaptorQ} majorly improve encoding
and decoding time, as well as decoding probability, as compared to \emph{\ac{LT} coding}.
\emph{RaptorQ} codes have been specified in RFC~6330~\cite{rfc6330} and represent the most advanced
version of fountain coding.

\begin{figure}[t]
	\centering
	\includegraphics[width=\columnwidth]{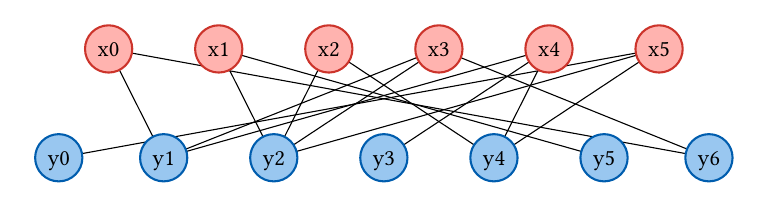}
	\caption{Bipartite graph with red source symbols and blue encoding symbols.}%
	\label{fig:fountain}
\end{figure}

\emph{\ac{LT} coding} can be represented as a bipartite graph, where encoding symbols
represent linear combinations of source symbols with XOR operations, as shown in
Figure~\ref{fig:fountain}. \ac{LT} coding is inherently probabilistic, as the
distribution of edges in this linear system is random.
Qualcomm explains~\cite{qualcommRaptorQ2010} the inner workings of edge
selection and the specific encoding and decoding procedures.

In summary, encoding symbols are formed by XOR operations, which correspond to
addition in $\mathrm{GF}(2)$ but can be generalized to
$\mathrm{GF}(q)$~\cite{qualcommRaptorQ2010}.
For $k$ source symbols, each encoding symbol corresponds to a coding vector $c_i = (c_{i,1}, c_{i,2}, \dots, c_{i,k}) \in \mathrm{GF}(q)^k$,
a vector space of dimension $k$ containing $q^k$ elements.
Recovering the source symbols therefore requires $k$ linearly independent coding vectors.
For uniformly random coding vectors, the probability that $k + h$ received symbols suffice is approximately $1 - q^{-(h+1)}$,
an accurate estimate for large $q$.
Each additional overhead symbol $h$ therefore reduces the failure probability.

As encoding symbols are mere linear combinations of the source symbols, an encoder can generate as many as needed,
unlike Reed-Solomon codes, whose number of distinct code symbols is bounded by the field size.
\end{description}

\subsection{QUIC}%
\label{sec:bg_quic}
Historically, most encrypted web traffic used SSL/TLS over TCP\@. This comes with
disadvantages, one of them being additional \acp{RTT} for first the transport and
only then the cryptography~\cite{quic}. In response, QUIC was designed with the idea to
integrate both handshakes into one protocol. It gives reliable, ordered,
congestion-controlled delivery with its own loss detection~\cite{quic_recovery} for
the cryptographic handshake as well as the payload connections.

QUIC was designed with the combination with TLS~1.3 in mind and to date only this
one is standardized~\cite{quic_tls}. However, QUIC itself only defines a slot for the
cryptographic protocol without mandating a specific one. The messages of QUIC's
handshaking phase travel in \emph{CRYPTO} frames which are structured
into three levels: \emph{Initial}, \emph{Handshake}, and \emph{1-RTT}. Each level has
one byte stream which QUIC splits across packets. All packets are encrypted, although the
key for the initial level's frames is derived from the connection ID, which is sent in cleartext, thus providing
obfuscation only. After finishing the initial level, the parties have established a fresh secret key that is then used for
protecting the following messages. The handshake level completes the handshake and the 1-RTT
level allows exchanging further CRYPTO frames or application data inside
\emph{STREAM} frames.

QUIC requires the IP network to carry at least 1200-byte UDP datagrams and enforces
each datagram carrying an initial packet to be padded to that size if necessary. This
impedes \ac{DoS} amplification attacks as it increases the cost for the client.
Additionally, a responder must not send more than three times the number of bytes it has
received from an address until that address's return-routability is proven. Under load, a responder may also frontload
return-routability verification by answering a connection attempt with a \emph{Retry}
message, similar to IKEv2's cookie mechanism. Furthermore, QUIC starts with
a small congestion window. Afterwards, it adjusts it to the observed network path's
capabilities. Large messages in the initial level may, therefore, take more \acp{RTT}
than in later handshaking stages.

QUIC identifies connections by connection IDs instead of the used addresses. This
gives roaming capabilitiesi, meaning adjusting to new addresses after validating the
new path. A connection's transport parameters, like the ID or flow control limits,
usually ride a respective TLS extension. Further features of QUIC entail ordering
application data into streams, reliably transmitting and multiplexing those streams,
and 0-\ac{RTT} session resumption. Those features allow QUIC's integration into the
HTTP/3 stack but do not affect any QUIC-IKEv2 interactions described in
Section~\ref{sec:quike}.

\subsection{Related Work}%
\label{sec:rel_work}
There is exhaustive research about \ac{PQC} migration and cryptographic agility.
Näther et al.~\cite{slr,ca} surveyed both areas and provide an overview of the
current literature. Yet, the research about the \ac{PQC} migration of secure
protocols is more fractured, tailored toward the specific protocol and committee that
maintains it.

One prominent protocol to be migrated is TLS 1.3~\cite{tls1.3}. Compared to IKEv2,
it sits on top of TCP which automatically remedies IKEv2's transport challenges.
Classic McEliece, however, also does not fit into the header fields.
Adding \ac{PQC} authentication in TLS is rather
straightforward, as the \ac{PQC} signature schemes behave similarly to their
classical counterparts except for their larger size, which is manageable with a TCP
underlay. Sikeridis et al.~\cite{tls13_auth_perf} assess \ac{PQC} authentication's
performance impact on TLS 1.3 and Paquin et al.~\cite{tls13_perf} inspect the
performance of \ac{PQC} on TLS 1.3 overall.

SSH uses TCP transport as well and chose similar hybrid key agreement schemes as
TLS\@. Sowa et al.~\cite{pqc_adoption} concluded that TLS and SSH were the quickest
to adopt (not necessarily standardized)
\ac{PQC} which we assume comes from them not having to solve as many transport
challenges scaling with key/ciphertext size.

In contrast, WireGuard~\cite{wireguard} also relies on UDP which introduces the same
challenges regarding the \ac{PMTU} and middleboxes as in IKEv2.
It already has a \ac{PSK} interface to be able to integrate keys exchanged
out-of-band. This, however, is more of a band-aid than a solution. Its full
\ac{PQC} version~\cite{pq_wireguard} solves this by selecting only \acp{KEM} that keep the in-band data small.
WireGuard never allowed any algorithm selection and, hence, can just pick a
small enough combination. It simply does not address needs for more
conservative algorithm choices.
WireGuard~\cite{wireguard} and PQ-WireGuard~\cite{pq_wireguard} do not use signatures
for authentication, in line with the Noise framework~\cite{noise} and its
post-quantum variant~\cite{pq_noise} or Rosenpass~\cite{Rosenpass}.
There, the authentication comes from the key
agreement itself. For this, the parties store a static key share~(\ac{DH}) / key
pair~(\ac{KEM}) which means that after key agreement, only the actual peer is able to
obtain the shared key, implicitly authenticating them (assuming no secret key got
lost). They combine this with an ephemeral key agreement for forward secrecy. That
authentication structure saves a second, separate cryptographic migration for
signatures. Yet, for the \acp{KEM}, it faces the same size question as IKEv2.
Schwabe et al.~\cite{kemtls} also proposed a modified version of TLS, incorporating
\ac{KEM} authentication. Yet, this solution struggles with adoption, mostly
as it represents a highly intrusive change with far-reaching consequences for current
deployments.

IKEv2, despite its design decisions of classic UDP and high adjustability to
different requirements, still took similar paths. One of the first proposals was to
add a special \ac{PSK} interface~\cite{ppk} to allow out-of-band exchange of keys,
similar to WireGuard's and IKEv1's~\cite{ikev1} \ac{PSK} mechanism. Similarly,
especially for the largest \acp{KEM}, the IETF working group discussed applying the
\emph{Hash-and-URL} mechanism, already standardized for large certificate
chains~\cite{ikev2}, to key exchange payloads. There, the peers only send a hyperlink
to the public key and a verification hash. Yet, merely moving the problem out of the
protocol does not solve it, which is also why the IETF discarded the idea
again~\cite{beyond_64k}. For \ac{PQC} authentication, there is one draft
defining the integration of \ac{PQC} signatures into IKEv2~\cite{ikev2_sig_auth}. As
they do not need to ride the first, non-fragmentable IKE\_SA\_INIT exchange, the
draft does not need to fundamentally alter the protocol.  Alternatively, there is also a proposal for \ac{KEM}-based
authentication in IKEv2~\cite{ike_kem_auth} which especially makes use of an
ML-KEM ciphertext being significantly smaller than an ML-DSA
signature. However, this proposal is even more dependent on a
fast and stable way to transmit \ac{PQC} \acp{KEM} in the first place. 

Another set of
proposals is concerned with hybrid key exchanges in the sense of combining
cryptography with \ac{QKD}, i.e., using quantum physics to exchange
secrets~\cite{muckle,muckle_plus}. These proposals, however, come with their own considerations
and are outside of the scope of this study.
Furthermore, there are several works analyzing specific approaches to integrating
\ac{PQC} key agreement into IKEv2~\cite{real_ipsec,pqc_macsec,ipsec_pqc_perf,ikev2_formal} or in
specific networking contexts, like 6G~\cite{ike_6g}, automotive~\cite{ike_automotive}
or satellite connections~\cite{ike_satellites}. Yet, none of them provides an
evaluation of the whole breadth of the current proposals under equal and exhaustive
circumstances.

Whether new approaches regarding protocol adjustments get standardized or not
ultimately depends on the respective IETF working group. Failing to meet their
requirements leads to even a theoretically sound proposal remaining theory. Eliot
Lear~\cite{rfc7305} and Welzl et al.~\cite{failedIETF} investigated the decisive
factors for success or failure at the IETF\@. Their results show that practicality in
terms of implementation and deployment is similarly important as measurable
properties, like performance or theoretical security properties. This mandates this
work to emphasize a holistic discussion about not only measured results but also empirical
results regarding implementation and deployment.

\section{Methodology}%
\label{sec:methodology}
In this section, we detail our research questions including their meaning and how we
aim to solve them. Further, we outline how we performed the systematic search for
proposals regarding \ac{PQC} in IKEv2, following the PRISMA 2020
guidelines~\cite{Prisma2020}. Finally, we explain the measurement testbed and
methodology used to evaluate the approaches' performance and network resilience.

\subsection{Research Questions}%
\label{sec:rqs}
In this work, we aim to address the following research questions:
\begin{enumerate}[label=\textbf{RQ\arabic*:}, ref=RQ\arabic*, leftmargin=2em, labelindent=0pt, itemindent=*]
	\item\label{rq:ietf_proposals} \emph{Which protocol adjustments for transmitting or optimizing the
		transmission of \ac{PQC} KEM-keys in IKEv2 have been proposed at the IETF\@?}

	 \emph{RQ1} examines all proposals for adding potentially large \acp{KEM} to IKEv2 via
		protocol adjustments. We focus on the IETF's corpus as that is the place where
		the proposals are discussed and there is no significant hurdle to submitting a draft
		and be heard by the committee. Also, we focus on genuine protocol adjustments and
		not implementation optimizations that one party can unilaterally just do.

	\item\label{rq:new_proposals} \emph{Which alternative protocol adjustments can further improve
		sending \ac{PQC} \acp{KEM} over IKEv2?}

	  \emph{RQ2} asks for novel approaches that represent alternatives to the proposals
		found in \emph{RQ1}. We address this research question by designing three new
		approaches, one using \ac{FEC}, one utilizing QUIC features~(QUIKE), and QUIKE
		with a composite key exchange.

	\item\label{rq:performance} \emph{Which performance implications do the different transmission
		techniques have, especially considering varying networking conditions?}

	 \emph{RQ3} investigates the performance and resilience to network faults of the
		approaches of \emph{RQ1} and \emph{RQ2}. For that we implemented all approaches
		on the same open-source codebase and measured them under the same conditions.

	\item\label{rq:implications} \emph{What are the implications, considering both the performance as
		well as the implementation costs, for adopting the strategies to real-world VPN
		implementations and deployments?}

		As discussed in Section~\ref{sec:rel_work}, the practicality in terms of both
		implementation and deployment is decisive for any approach to be considered
		for standardization. \emph{RQ4} takes the results from \emph{RQ3}, adds the factors regarding
		the approaches' implementation, and discusses all those implications, yielding a
		holistic view on the approaches.
\end{enumerate}
\subsection{Systematic IETF Proposal Search}%
\label{sec:sys_review}
The basis for our systematic search of the IETF's IPsec working group's document
body for \ac{PQC} integration proposals into IKEv2 consists of the PRISMA 2020
guidelines~\cite{Prisma2020}. They were originally designed for a full systematic review of
academic literature. We limited the scope to the respective IETF documents and
transferred the guidelines to the new scope.

The objective is to find all IETF proposals for protocol adjustments to facilitate
\ac{PQC} \ac{KEM} integration into IKEv2. We see the limitation as sensible, as the
IPsec working group is where those are discussed and there is no significant barrier
to submitting a respective draft. Hence, we assume any promising candidate potentially published elsewhere is also part of
the IETF's discussion. Adjustments to adjacent protocols, on the other hand, are
discussed in Section~\ref{sec:rel_work}.

For the search, we mirrored a snapshot of the Internet-Draft archive, including
active and expired revisions, and RFC series via
rsync\footnote{rsync.ietf.org::(id-archive)|(rfc); Date: 27.08.2026}. We scanned
the resulting $189\,328$ documents for the keywords in
Table~\ref{tab:keywords}, with inclusion requiring at least one hit in each
category. Matching is case-insensitive, flattens whitespaces, and does not
discriminate between hyphens and spaces. The keywords are comparatively broad and hence
prioritize recall over precision. This is a trade-off to minimize the potential for
missed proposals. Finally, the automatic selection process removed older revisions of
one document, i.e., documents with the same name but a lower version index at the end of
their name. For reproducibility, the replication package contains the script for screening,
the snapshot timestamp, and the document counts at each stage.
\begin{table}[!tb]
	\caption{The screening keywords for the IETF document search.}%
\label{tab:keywords}
\centering
\begin{tabularx}{\columnwidth}{lX}
\toprule
\textbf{Category} & \textbf{Keywords} \\
\midrule
Protocol & ikev2, ike\_sa\_init, ike\_auth, ike\_intermediate, rfc 7296,
	ipsec \\
\ac{KEM} & post-quantum, quantum-resistant, kem, ml-kem, kyber, mceliece,
	frodo, sntrup, hybrid key exchange \\
Size & fragment, large, mtu, packet size, datagram, retransmi, congestion,
	amplificat, 65535, reliab, transport, compress, hash-and-url,
	out-of-band \\
\bottomrule
\end{tabularx}
\end{table}

\begin{table}[!t]
\caption{Inclusion and exclusion criteria for the manual screening.}%
\label{tab:criteria}
\centering
\begin{tabularx}{\columnwidth}{llX}
\toprule
\textbf{Code} & \textbf{Type} & \textbf{Criterion} \\
\midrule
I1 & inclusion & Proposals within official IETF documents \\
I2 & inclusion & Proposals for new protocol adjustments facilitating or optimizing the exchange of larger key agreement payloads over IKEv2 \\
\midrule
D1 & dedupl. & Duplicates of another in-scope document or mechanism \\
D2 & dedupl. & Older versions of another in-scope document \\
\midrule
E1 & exclusion & Non-IETF gray literature, i.e., blog posts, websites, RSS feeds, videos, podcasts and webinars \\
E2 & exclusion & Literature that is not available \\
E3 & exclusion & Literature in any language other than English \\
E4 & exclusion & Proposals that focus on QKD rather than PQC \\
E5 & exclusion & Proposals that focus on authentication rather than key agreement \\
E6 & exclusion & Proposals that rely on out-of-band key transmittal \\
E7 & exclusion & Proposals that rely on middleboxes supporting IP fragmentation \\
E8 & exclusion & Mere assessment or analysis of existing proposals \\
E9 & exclusion & Prerequisites for all other proposals, affecting them evenly (e.g., algorithm naming, larger header fields) \\
\bottomrule
\end{tabularx}
\end{table}

One author screened the entire resulting document set in two passes. In the first
pass, they evaluated each document and applied the criteria listed in
Table~\ref{tab:criteria}. A document is included if every
inclusion criterion matches and no deduplication or exclusion criterion does. They first checked the documents'
scope~(inclusion), then removed duplicates, and finally applied specific exclusion criteria.
They noted the first criterion leading to exclusion and did not evaluate the
remaining ones. A second pass then re-read the included documents and
those excluded as prerequisites. Each document proposing more than one mechanism was
split into one entry per mechanism. We evaluated those individually against the same
criteria, and, if excluded, again noted the first failing one.
Here, the inclusion criteria reflect \emph{\ref{rq:ietf_proposals}} and deduplication and criteria E1
to E3 are literature review standards. Exclusion criteria E4 and E5 remove proposals
for adjacent problems, E6 those that relocate the problem instead of solving it
(analogous to the Hash-and-URL mechanism discarded in
Section~\ref{sec:rel_work}), E7 is meant to catch impractical ones~\cite{ike_frag},
and E8 drops literature proposing nothing new. E9 excludes those mechanisms from the
comparison in \emph{\ref{rq:performance}} that affect all other proposals equally, and therefore
do not need a standalone evaluation.

To increase recall, we added a backward snowballing check, adding all IETF documents
referenced by the seed set and by the documents proposing included mechanisms. The
snowballing also excludes academic literature. Nevertheless, we validated a sample of
referenced academic literature, none of which introduced any new mechanism but was
background regarding \ac{PQC} and IKEv2, further justifying our limitation to IETF
documents.

\subsection{Measurement Approach}%
\label{sec:measurement_setup}
\begin{table}[!tb]
	\caption{Grid of evaluated \acp{KEM} and networking conditions. Every combination
	of the different factors is tested.}%
\label{tab:grid}
\centering
\begin{tabularx}{\columnwidth}{l>{\raggedright\arraybackslash}X}
\toprule
\textbf{Factor} & \textbf{Invariants} \\
\midrule
\ac{KEM}						& ML-KEM-1024, FrodoKEM-1344,
											\mbox{Classic McEliece 8192128} \\
One-way delay (ms)  & $50\pm5$, $100\pm10$, $200\pm20$ \\
Bandwidth (Mbit/s)  & 1, 10, 100 \\
Packet loss (\%)    & 0, 0.1, 0.5, 2, 5, 10 \\
\midrule
Reorder (\%)				& 5 (fixed condition) \\
\bottomrule
\end{tabularx}
\end{table}

To answer \emph{\ref{rq:performance}}, we first implemented all five approaches found in \emph{\ref{rq:ietf_proposals}}
together with our own three from \emph{\ref{rq:new_proposals}} and then created a dedicated hardware
testbed, in which
we can compare the different implementations for \ac{PQC} \ac{KEM} integration under
various networking conditions. Authentication is not in the scope of this study.
Thus, we use static \ac{PSK} authentication within the test setup. Our evaluation
uses three \acp{KEM}: \mbox{ML-KEM}~\cite{mlkem}, FrodoKEM~\cite{frodo}, and Classic
McEliece~\cite{mceliece}, all in their largest parameter set classified as
Level 5 by \ac{NIST}~\cite{nist_pqc_cfp}. Table~\ref{tab:grid} lists the evaluated
\acp{KEM} and network conditions. Reordering is always to be expected when communicating over the
Internet~\cite{reordering}, so we kept that as a constant $5\%$. The delay's jitter also does
not further increase the reordering as it is correlated by $90\%$. For each
combination of factors, we measured the following metrics:
\begin{itemize}
	\item The handshake completion rate
	\item The handshake completion time
	\item The number of packets sent per direction
\end{itemize}

The testbed used to perform the evaluation consists of two directly connected,
bare-metal Linux hosts with a 7.0.0-30-generic kernel on an
Intel Xeon E-2236 CPU at $3.40$\,GHz with $2\times8$\,GiB DDR4 RAM.
The TCP arms use the kernel's default CUBIC congestion control and all arms
link OpenSSL~$3.5.1$.
This setup grants a stable and fast networking environment.
To prevent inflated packet counts on the interfaces, we set static ARP entries on
both hosts. Also, both machines use a separate network connection for management and
test orchestration. The link speed is $10$\,Gbit/s and the testbed's unimpaired
\ac{RTT} stays below $1$\,ms, negligible against every combination of
Table~\ref{tab:grid}.

The network emulation tool \emph{netem}~\cite{netem} applied the networking conditions
to both hosts' egress interface. Hence, the one-way delays $d$ of
Table~\ref{tab:grid} apply once per direction causing an \ac{RTT} of about
$2d$. With hardware offloading (GRO/TSO/GSO/LRO) enabled, the network stack hands
netem segments exceeding the \ac{MTU}\@. The rate shaper mishandles them,
empirically collapsing TCP results while leaving UDP-based ones intact. To
keep a fair comparison, we turned off all hardware acceleration. The test setup uses
random, uncorrelated loss~\cite{netem}. Its lower variance increases comparability but
depicts real-world loss events less accurately~\cite{loss_burst}.

Testing starts with the controller, which runs on the machine of
the initiator, copying each binary and its necessary libraries to the
responder, making sure both machines test the exact same implementation. Then, it
randomly schedules the order of different configurations to be tested and the, again
random, order of implementations for each configuration. Both orders are seeded and
therefore reproducible. The test is a \emph{systemd}
service which restarts automatically on reboot or other disruptions. After such an
event, the testing continues where it stopped and keeps the initially determined
ordering.

At the beginning of every configuration, the controller first re-asserts both
networking configurations, including addresses, ARP entries, offloading, and netem
emulation. Then it creates the respective IKEv2 configurations, including the fixed
\ac{PSK} authentication. With those preconditions met, it starts the IKEv2 process on
the responder side which runs until testing has finished for this configuration. To
prevent the responder from sending keep-alives for its already completed handshake,
inflating packet counts, it has dead peer detection turned off. Finally, it runs the
IKEv2 daemon on the initiator side, freshly restarting it for each handshake. The
controller attaches to the initiator's log file, timestamping the exact moment it
sends its initial packet and the moment it installs its first child \ac{SA}. This
interval represents the handshake completion time. One run finishes either with this event,
which triggers the restart of the initiator, or alternatively any logged failure to
perform the handshake, reducing the handshake completion rate. This can be, e.g., due to the maximum number of retries being reached, or
aborting the run after $300$\,s, which is also the maximum exchange timeout of the IKEv2 implementation we used. After each handshake, the controller
extracts the pre-drop packet count of each netem queue, yielding the packets sent for
each direction.

\begin{algorithm}[!t]
\caption{The algorithm running one test configuration and deciding when the number of
	handshakes $n$ is sufficient. $h_p$ and $U$ are the half-width and the upper bound
	of the Wilson interval for the completion rate, $h_\theta$ the bootstrap half-width
  of the median completion time.}%
\label{alg:stopping}
\begin{algorithmic}[1]
\REQUIRE batch size $b{=}10$, $\nmin{=}30$, $\nmax{=}200$
\REQUIRE targets $\epsilon_p{=}0.03$, $\epsilon_\theta{=}0.10$, decisive bound $u{=}0.20$
\STATE run and discard one warmup handshake
	\STATE $s \gets 0$ \COMMENT{consecutive-check counter}
	\STATE $n \gets 0$  \COMMENT{number of performed handshakes}
\REPEAT
	\STATE run $b$ handshakes
	\STATE $n \gets n+b$,\quad $k \gets$ number of successful handshakes
	\STATE $\hat{\theta} \gets$ median of the completion time \COMMENT{failures counted as
	$\infty$, undefined if $k \leq n/2$}
	\IF{$U(k,n) \leq u$}
		\STATE $m \gets \textnormal{\textbf{true}}$ \COMMENT{failing configuration $\rightarrow$ abort}
	\ELSE
		\STATE $m \gets \big(h_p(k,n) \leq \epsilon_p\big) \wedge
		        \big(\hat{\theta}\text{ undefined} \vee h_\theta \leq \epsilon_\theta \hat{\theta}\big)$
	\ENDIF
	\IF{$n < \nmin$}
	\STATE $m \gets \textnormal{\textbf{false}}$ \COMMENT{only abort after at least $\nmin$ tries}
	\ENDIF
	\IF{$m$}
		\STATE $s \gets s+1$
	\ELSE
		\STATE $s \gets 0$
	\ENDIF
\UNTIL{$(m \wedge s \geq 2) \vee (n \geq \nmax)$}
\RETURN
\end{algorithmic}
\end{algorithm}

The number of handshake runs per configuration and approach $n$ is not fixed.
Instead, we focus on the test configurations with high variance, granting them more
runs and spending less time on the decisive ones. So, $n$ is theoretically in the interval of $40 \le
n \le 200$. Its specific number depends on the confidence intervals of both the
success rate and the completion time's median. In almost all configurations, the success rate is usually close to
$1$. There, a classical confidence interval lacks
coverage and we instead use the Wilson score interval. The completion time does not
follow a binomial model, hence its confidence interval comes from a percentile
bootstrap. Algorithm~\ref{alg:stopping} shows how $n$ is exactly determined. After
one discarded warmup handshake, it runs batches of size $b = 10$. After each batch,
it verifies that at least $\nmin = 30$ runs were conducted. Then, the completion
criteria $m$ are satisfied if either the configuration is failing decisively, or both
confidence intervals are small enough. As soon as $m$ is set for two consecutive
batches, or as soon as $n$ reaches $\nmax$, testing finishes for this
configuration. Note that because $m$ has to be satisfied for two consecutive batches after finishing at least \nmin runs, we have at least 40 runs, even though we set $\nmin = 30$.

The test and analysis scripts, as well as the complete results, are part of the
replication package. These allow reproducing the same test setup. The exact testing
order of the different configurations is seeded and thereby also reproducible. Within
one configuration all approaches run back-to-back, mitigating testbed drift. However,
in case outside disturbances occurred, we pruned the respective results and reran
those specific configurations, taking them out of the initially planned order. Also,
netem cannot be seeded, which gives statistical reproducibility only. Still,
for our values of $n$, the measurements are expected to yield the same statistical
results.

\section{\acs{PQC} in IKEv2}%
\label{sec:ikev2_pqc}
This section answers \emph{\ref{rq:ietf_proposals}} and \emph{\ref{rq:new_proposals}}. For \emph{\ref{rq:ietf_proposals}} we present the results
of the systematic search outlined in Section~\ref{sec:sys_review} and explain
the found approaches in detail. For
\emph{\ref{rq:new_proposals}} we outline our novel approaches and how they differ from the existing
ones found in \emph{\ref{rq:ietf_proposals}}.

\subsection{Results of the Systematic Search}%
\label{sec:search_results}
\begin{table}[!tb]
	\caption{The screening results described in Section~\ref{sec:sys_review} and for
		the criteria as defined in Table~\ref{tab:criteria}. The documents come from the
		first and the mechanisms from the second pass.}%
\label{tab:screening}
\centering
\begin{tabular}{llcc}
\toprule
\textbf{Code} & \textbf{Outcome} & \textbf{Documents} & \textbf{Mechanisms} \\
\midrule
	& Screened                   & 163 & 168 \\
\midrule
I1	& Not an IETF document    & 0 & 0 \\
I2	& Out of scope            & 127 & 130 \\
\midrule
 & In scope											& 36 & 38 \\
\midrule
D1	& Duplicate               & 0 & 2 \\
D2	& Superseded              & 9 & 9 \\
\midrule
 & Individual in scope				& 27 & 27	\\
\midrule
E1	& Gray literature         & 0 & 0 \\
E2	& Not available           & 0 & 0 \\
E3	& Not English             & 0 & 0 \\
E4	& \acs{QKD} rather than \acs{PQC} & 1 & 1 \\
E5	& Authentication          & 11 & 11 \\
E6	& Out-of-band key distribution         & 2 & 2 \\
E7	& Relies on middleboxes   & 0 & 0 \\
E8	& No new mechanism        & 0 & 0 \\
E9	& Prerequisite / even effect            & 8 & 8 \\
\midrule
	& Included                & 5 & 5 \\
\bottomrule
\end{tabular}
\end{table}

\begin{table}[!tb]
	\caption{The mechanisms addressing large \acp{KEM} in IKEv2 found by the
	systematic search. Status is the standardization state at the time of the
	snapshot.}%
\label{tab:mechanisms}
\centering
\begin{tabularx}{\columnwidth}{Xll}
\toprule
\textbf{Mechanism} & \textbf{Status} & \textbf{Doc.} \\
\midrule
\multicolumn{3}{l}{\textit{Included proposals}} \\
Additional key exchanges  & RFC        & \cite{rfc9370} \\
Separate IKE and ESP transports  & WG         & \cite{reliable_transport} \\
Per-fragment acknowledgment      & individual & \cite{fragment_ack} \\
Selective retransmission         & individual & \cite{selective_retransmission} \\
Composite key exchange~(TCP)           & individual & \cite{hybrid_reliable} \\
\midrule
\multicolumn{3}{l}{\textit{Excluded as prerequisites / even effect (E9)}} \\
32-bit payload length            & individual & \cite{big_payload} \\
Extended payload header          & individual & \cite{extended_pld} \\
Split across adjacent payloads           & individual & \cite{beyond_64k} \\
Denser payload encoding          & individual & \cite{compact_pld} \\
\addlinespace
\multicolumn{3}{l}{\textit{Excluded as codepoint registrations (E9)}} \\
ML-KEM                           & WG         & \cite{mlkem_ikev2} \\
FrodoKEM                         & WG         & \cite{frodo_ikev2} \\
Classic McEliece                 & individual & \cite{mceliece_ikev2} \\
NTRU                             & individual & \cite{ntru_ikev2} \\
\bottomrule
\end{tabularx}
\end{table}

The automatic search for the keywords from Table~\ref{tab:keywords} yielded $649$
documents. After the automatic removal of older revisions, $163$ documents were left
for manual screening~(Table~\ref{tab:screening}). After deduplication, they offered $27$ mechanisms related to
\ac{PQC} in IKEv2. Almost half~($11$) of the mechanisms were concerned with authentication,
one with \ac{QKD}, and two required out-of-band key distribution, leaving $13$
actually addressing large \ac{PQC} \acp{KEM} in IKEv2. We classified five of those as
genuine new mechanisms for the comparison in \emph{\ref{rq:performance}} and eight as either technical
prerequisites or optimizations that affect all branches similarly, as shown in
Table~\ref{tab:mechanisms}. Only additional key exchanges has completed
standardization.
The drafts for separate transport layers for IKE and ESP and the codepoint
registration for ML-KEM and FrodoKEM are adopted by the working group. Every other
mechanism is, at the date of the snapshot, an individual draft, highlighting
the still active discussion at the IETF\@. We implemented all
included mechanisms and the 32-bit payload length draft to be able to encode a
Classic McEliece public key for all other mechanisms. We opted for that one for
mere implementation reasons, as none of the three proposals addressing that
limit severely affects the transport. Finally, we added codepoints for all
\acp{KEM} that we tested to our implementations.

\subsubsection{\acs{PQC} Prerequisites \& Optimizations in IKEv2}%
\label{sec:prereq}
A key exchange payload holding \ac{DH} key shares or \ac{KEM} public keys and
ciphertexts has a header with a length field of $16$ bits. This caps a payload,
including its header, at $64$\,KiB. A Classic McEliece public key of $\sim1.4$\,MB
clearly exceeds this length. The first three proposals excluded by \emph{E9} address
this gap using different mechanisms. The 32-bit payload length~\cite{big_payload}
uses one of the header's reserved bits to indicate that the following payload has a
double sized payload length field of $32$ bits. Using reserved bits is always
controversial but we saw it as the most straightforward implementation. Hence, we
implemented this as the prerequisite for all the other proposals to enable their
Classic McEliece test case. The other options are defining a new, extended
header~\cite{extended_pld} or splitting one actual payload over several smaller,
adjacent ones~\cite{beyond_64k}. The decision about standardizing one of those
solutions is up to the working group and not considered further in this study.

The fourth mechanism marked as \emph{E9} allows to encode the payload header more
densely~\cite{compact_pld}. It is meant for resource-constrained environments where
saving single bytes increases the devices' availability. The overhead of payload
headers is negligible compared to the tested \ac{PQC} \acp{KEM} and the reduction is
flat and does not scale with payload length. Furthermore, even with that reduction
each proposal with every \ac{KEM} surpasses the fragmentation threshold, keeping the
same exchange structure. Thus, we excluded it from further investigation during this
study.

\begin{figure}
	\centering
	\providecommand{\imagedir}{./}
\input{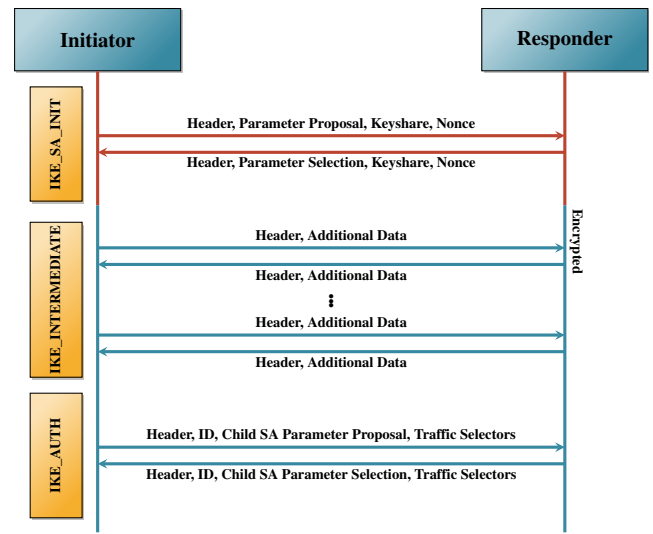}

\begin{tikzpicture}
	\def\width{450 pt}
	\def\rowsep{150}		
	\def\plainrun{129}		
	\def\encrun{315}		

	\node (init) [blue_box] {\huge\textbf{Initiator}};
	\node (resp) [blue_box] at ([xshift=\width] init)
		{\huge\textbf{Responder}};

	\draw[lifeline, draw=red] (init.south) -> ++(0,-\plainrun pt) coordinate(i1);
	\draw[lifeline] (i1) -> ++(0,-\encrun pt);
	\draw[lifeline, draw=red] (resp.south) -> ++(0,-\plainrun pt) coordinate(r1);
	\draw[lifeline] (r1) -> ++(0,-\encrun pt) node[ sloped, pos=0.11,
		above] {\bfseries\Large Encrypted};

	\coordinate (sa_init) at ([yshift=-70]init.south);
	\node[exchange] at ([xshift=-40]sa_init) {\bfseries\Large IKE\_SA\_INIT};

	\draw[arrow, draw=red] ([yshift=8]sa_init) -> ++(\width,0)
		node[pos=0.5, above] {\bfseries\Large Header, Parameter Proposal,
		  Keyshare, Nonce};
	\draw[arrow, stealth-, draw=red] ([yshift=-8]sa_init) -- ++(\width,0)
		node[pos=0.5, below] {\bfseries\Large Header, Parameter Selection,
		Keyshare, Nonce};

	\coordinate (intermediate) at ([yshift=-\rowsep] sa_init);
	\node[exchange] at ([xshift=-40]intermediate)
		{\bfseries\Large IKE\_INTERMEDIATE};

	\node at([shift={(\width/2pt,2pt)}]intermediate)
		{\bfseries\Huge \vdots};
	\draw[arrow] ([yshift=50]intermediate) -> ++(\width,0)
		node[pos=0.5, above] {\bfseries\Large Header, Additional Data};
	\draw[arrow, stealth-] ([yshift=34]intermediate) --
		++(\width,0) node[pos=0.5, below] {\bfseries\Large Header,
		Additional Data};
	\draw[arrow] ([yshift=-34]intermediate) -> ++(\width,0)
		node[pos=0.5, above] {\bfseries\Large Header, Additional Data};
	\draw[arrow, stealth-] ([yshift=-50]intermediate) --
		++(\width,0) node[pos=0.5, below] {\bfseries\Large Header,
		Additional Data};

	\coordinate (ike_auth) at ([yshift=-\rowsep]intermediate);
	\node[exchange, minimum width=120] at ([xshift=-40]ike_auth)
		{\bfseries\Large IKE\_AUTH};

	\draw[arrow] ([yshift=8]ike_auth) -> ++(\width,0)
		node[pos=0.5, above] {\bfseries\Large Header, ID, Child SA
		Parameter Proposal, Traffic Selectors};
	\draw[arrow, stealth-] ([yshift=-8]ike_auth) --
		++(\width,0) node[pos=0.5, below] {\bfseries\Large Header, ID,
		Child SA Parameter Selection, Traffic Selectors};

\end{tikzpicture}
	\caption{The IKEv2 message flow including the IKE\_INTERMEDIATE exchange.}%
	\label{fig:intermediate}
\end{figure}
One prerequisite that is designed with \ac{PQC} \acp{KEM} in mind is the intermediate
exchange~(\emph{IKE\_INTERMEDIATE}) defined in RFC~9242~\cite{rfc9242}. It does not
define its interaction with \ac{PQC} itself which excludes it from the survey but
most included mechanisms rely on it as a basis. The intermediate exchange is a new
exchange type in IKEv2 that sits between the initial and the authentication exchange.
This position allows it to be fragmented, hence, also to carry payloads larger than
the \ac{MTU}. Figure~\ref{fig:intermediate} shows the resulting exchange structure of
IKEv2. There can be seven intermediate exchanges in one handshake and each exchange
is encrypted with the latest round of keying material, i.e., the first uses the key
exchanged in the initial exchange. Each of these intermediate rounds costs one
additional \ac{RTT} in terms of performance and speed. Despite the encryption the
peers are not yet authenticated, hence, we have to expect any peer to act maliciously
during the intermediate exchange.

The final prerequisite is the codepoint registrations. Before any IKEv2
implementation can negotiate a \ac{PQC} \ac{KEM}, it needs a transform identifier.
There exists a draft for all three \acp{KEM} we test, as well as one for NTRU which
we do not test, but none of them has an assigned identifier yet. Therefore, our
implementations use identifiers from the private-use range which are then not
interoperable with other implementations but prove the functionality for our tests in
\emph{\ref{rq:performance}}.

\subsubsection{IETF \acs{PQC} Proposals in IKEv2}%
\label{sec:proposals}
The systematic search found the five approaches we included in our evaluation. The
first is the already standardized
RFC~9370~\cite{rfc9370} allowing the \acf{ADDKE} shown in
Figure~\ref{fig:intermediate}. It directly builds on the intermediate exchanges,
allowing one additional key agreement per intermediate exchange, up to seven
additional key agreements in total. After each key agreement, there is a new key derivation
and the next exchange uses this new key. Hence, an attacker needs to break the
different keys in order to be able to decrypt the next key agreement packets.
IKEv2's native fragmentation~\cite{ike_frag} can apply to intermediate exchanges,
enabling the transmission of large \acp{KEM}. However,
this approach is complex to implement due to the interaction of fragmentation and the
handshake signature calculation and IKEv2 fragmentation struggles with large messages
over constrained networks~\cite{real_ipsec}. This comes from the retransmission logic
described in Section~\ref{sec:bg_ikev2}.

There are two documents proposing optimizations to IKE fragmentation in the context
of \ac{ADDKE}. They try to improve
the fragmentation performance especially under lossy networking conditions by adding
transport-like features. The first is selective
retransmission~\cite{selective_retransmission}, referred to as \emph{status-ACK}
below. A peer that received some but not all
fragments of a message can report which ones are missing by sending a status message
under the same \ac{SA}. This status message carries the first and the last fragment's
number addressed by this packet. Further, it contains a bit vector over that range,
in which a $1$ marks a successfully received fragment and a $0$ one that is still
missing. Upon receiving such a status message, the sender retransmits the missing
fragments. This way, the approach reduces the time until loss is detected and at the
same time minimizes redundant data on the wire. The status message also contains its
own packet number, which the sender must increase with every status message within an
exchange. Although the draft mandates no receiver-side check, this number enables
detecting replayed status messages, which would otherwise trigger unwanted
retransmissions. This approach is not
negotiated but just assumed by a supporting peer. Its \ac{ICV} is calculated in a way
that leads to an unsupporting peer discarding and ignoring the messages.

The per-fragment acknowledgment~\cite{fragment_ack}, correspondingly
\emph{notify-ACK}, follows a similar idea. Yet,
instead of sending status messages, it sends distinct notify types for acknowledging
request and response fragments and instead of a bit vector it sends intervals of the
still missing packets. The intervals consist of a loss chunk's first and last
fragment
number. Upon receiving such a notify, a peer resends all fragments marked as missing.
The exact logic for when a peer should send these notifications is weakly defined and
up to the implementation. The draft suggests that a responder may wait for
retransmitted fragments as proof of packet loss before acknowledging, potentially
still taking long to successfully transmit all fragments~\cite{fragment_ack}. Unlike
status-ACK, support for this approach is negotiated during the initial
exchange. Further, the notify messages carry no message counter, so a receiver cannot
detect replayed notifies. Also, it has an open issue with reusing the \ac{AEAD}
nonce, which could break both confidentiality and authentication for that
\ac{SA}~\cite{fragment_ack}.

The last two approaches both rely on TCP instead of UDP, offloading the transport
challenges to a dedicated transport layer instead of selectively implementing
transport features inside of the security protocol. Tunneling IPsec including IKEv2
over TCP is already standardized in RFC~9329~\cite{rfc9329}. However, that approach
is motivated by middleboxes that may block the UDP-based connections. Therefore, it
tunnels both IKEv2 and ESP over one TCP connection. The payload data streams may also
use TCP which means, in the case of RFC~9329, we might tunnel one TCP connection in
another TCP connection. This can have unwanted side effects reducing the performance
and availability of the connections~\cite{tcp_meltdown}. The \ac{PQC}-motivated
drafts, on the other hand, propose only encapsulating the IKEv2 traffic in TCP while
the ESP traffic remains on a datagram transport. The first of these two approaches is the
separate transports draft~\cite{reliable_transport}. It tunnels IKEv2 in TCP,
removing the need for extra fragmentation and separate retransmission logic as TCP
handles all these transport challenges. The message flow remains as proposed in
RFC~9370~\cite{rfc9370}, shown in Figure~\ref{fig:intermediate}, with the only
difference being the additional TCP handshake. This can happen either after the
initial exchange or before the IKEv2 handshake. The former lets the peers negotiate
compatibility over UDP first, while the latter requires prior knowledge of the peer's
support as the handshake otherwise fails silently.

The last proposal found in the systematic proposal search is a derivative of the
separate transports approach. It uses the same transport layer assignment~(IKEv2 over
TCP, ESP over UDP) but instead of using intermediate exchanges it puts the \ac{PQC}
\ac{KEM} into one composite key exchange payload together with the classical
one~\cite{hybrid_reliable}. The message flow therefore falls back to the classic one
shown in Figure~\ref{fig:ikev2}, without any intermediate exchange. Afterwards, both
negotiated secret keys need to be combined in a suitable \ac{KDF}, also considering
the \ac{KEM} binding properties~\cite{holding_kems}. It removes additional \acp{RTT}
at the cost of a first message exceeding the \ac{MTU}. As the \ac{MTU} limitation
mainly comes from UDP-related considerations and secure \ac{KEM} combiners are
well-studied, the authors argue that this is a sensible trade-off and, hence,
adjustment to TCP-based IKEv2~\cite{hybrid_reliable}.

We implemented all five proposals identified in the IETF corpus to evaluate them
together with the following novel approaches in terms of performance and resilience.

\subsection{FEC in IKEv2}%
\label{sec:fec}
In the case of stabilizing \ac{VPN} connections, \ac{FEC} is used to absorb packet
loss. This requires a burst-resistant code as, to a decoder, a lost packet always
translates to a whole chunk of missing bytes, hence a burst. Realistically, packet
loss also happens in streams, therefore this burst may be very long. The best results
require a fitting \ac{FEC} style that is well-suited for this special loss pattern.
Afterwards we propose an implementation for erasure codings in IKEv2.

\subsubsection{Choosing an Erasure Coding}
Initially, hard resilience hints at a perfect use case for Reed-Solomon codes, but they exhibit two major
problems. First, their en- and decoding overhead is unattainable at the required throughput
speeds of a VPN concentrator, and second, encoding \ac{KEM} key material larger than $2 ^ {16}$
bytes is impossible without further complications. This leaves Reed-Solomon codes rather unattractive for this application.

Subsequently, there is the consideration between \ac{LDPC} and turbo codes, as both already compete in a
similar domain of wireless data transmission. Still, they differ substantially outside
the realm of random bit errors in channels. While \ac{LDPC} performs exceptionally well in low
noise environments~\cite{lei2024}, it usually struggles in burst scenarios~\cite{cui2026}.

The situation with turbo codes is more complicated. Shortly after their
publication in 1993, new research determined the effectiveness
of their interleaver in the case of burst channels~\cite{kang1999,sanchez2000}. In the context of wireless radio
transmission, such channel bursts usually occur in the form of fading channels, where a signal's
properties, such as amplitude or phase, change over time.
Xia~\cite{Xia2024} analyzed turbo codes from a signal processing viewpoint and concluded that turbo
codes perform better than conventional codes alone, but mentioned the interleavers' dispersion of
errors entering subsequent decoding steps, which may harm decoding success. As the potential
of dealing with burst errors originates from the power of the interleaver, the design itself is not
necessarily well-suited for packet loss-style bursts.

RaptorQ coding, on the other hand, treats all erasures uniformly, regardless
of temporal correlation. This results in missing packets never impacting others. This is a very
enticing characteristic, as it enables recovery to be much more predictable and removes structural
penalties which may be present in turbo decoding.
Because of this predictability, RaptorQ fountain coding allows for the computation of how many
encoding symbols are required for a highly successful decoding. This removes guesswork of how
much redundant data is required, as the amount is constant for RaptorQ regardless of loss
sequence, whereas for turbo codes it is not. Therefore RaptorQ fountain coding was chosen as
the best fitting candidate.

To encode data, RaptorQ splits up data into source blocks, which are then further split up
into source symbols~\cite{rfc6330}. RaptorQ encoding is applied to each source block individually, which
means that each block is able to reconstruct all data within itself. This is beneficial for streams
of data, but no tangible gain on singular transmissions emerged, so we chose to use a single
source block for each key material transmission.
RaptorQ, being a probabilistic encoding, requires a few additional packets to dramatically
decrease the chance of failing the decoding step, even if all sent packets \emph{are}
present.
To differentiate
between technically required overhead symbols and redundant extra symbols, the
implementation separates them into two
configurable values. Overhead symbols are an absolute additive, whereas extra symbols
superimpose a proportional increase. Later, the implementation calculates the total amount of fragments
and generates that many encoding symbols accordingly.
The remainder of this work uses the terms overhead and extra symbols. To depict either
or both redundant symbol kinds, we use the term additional or redundancy packet.

RaptorQ is covered by patents. For implementations of standards-track or experimental documents
outside wireless wide-area standards, the patent holder has declared that it will not assert them,
provided the implementation complies with RFC~6330~\cite{rfc6330,rfc6330_ipr}. Our implementation uses
RaptorQ unmodified, so our use falls within this declaration.

\subsubsection{Implementing FEC in IKEv2}
To integrate the presented \ac{FEC} into IKEv2, we propose the following
adjustments to IKEv2 fragmentation~\cite{ike_frag}:

\begin{description}
	\item[Negotiation.] Endpoints intending to use \ac{FEC}-enhanced fragmentation
		should send a notification with the notify payload type of
		IKEV2\_FEC\_SUPPORTED\@. Without any IANA ID assignment of this type, both peers
		must use a unique and so far unused identifier.

	\item[Using IKE Fragmentation.] If both parties expressed willingness, they must
		also use \ac{FEC} in order to prevent initial long timeouts.

	\item[Message Fragmentation.] Before fragmentation, the entire payload stream must
		be encoded with \ac{FEC}. The header stays untouched.
		This results in a list of encoding symbols, each
		prepended with a small header containing the unique symbol identifier. These
		symbols are then encrypted, authenticated, and then also prepended with the
		Encrypted Fragment payload header, as before. Selecting the fragment size must
		account for that additional header, which may cause more fragments to be
		transmitted, even without symbol-redundancy.

	\item[Receiving a Fragmented Message.] A receiver identifies each Encrypted Fragment
		payload as \ac{FEC}-encoded once both peers have signaled support. Unlike
		with classical fragmentation, decoding does not require all fragments. After
		decrypting a payload fragment and validating its authenticity, the content is
		buffered until reassembly.

		The number of fragments necessary for decoding depends on the used erasure coding
		and configuration. As soon as it is met, reassembly is triggered. Each fragment
		represents an encoded symbol, all of which are fed into the corresponding \ac{FEC}
		decoder. The decoded stream is identical to the result of classical IKEv2
		fragmentation. If decoding of a probabilistic \ac{FEC} fails, the assumed number
		of necessary packets was too small. Newly arriving packets retrigger the
		reassembly until it succeeds. If there are no additional packets, then the
		configuration of the erasure coding is incompatible with the lossy channel and
		must be adjusted.

	\item[Transport Considerations.] When using \ac{FEC}, single packets can be dropped
		up to the capability of the erasure coding and its used configuration. We expect
		a more resilient encoding configuration to be more time efficient than multiple
		elongated retransmits and timeouts.
\end{description}

\subsection{QUIKE --- QUIC using IKEv2}%
\label{sec:quike}
The \ac{FEC}-based approach adds redundancy to the UDP-based messages, decreasing the
likelihood of retransmissions being necessary. Most previous proposals, on the other
hand, aim at improving the retransmission mechanism, either by optimizing IKEv2's
native one or by mounting it on top of TCP as a reliable transport channel. One
drawback of TCP is that it adds another \ac{RTT} before the actual handshake. Instead
of TCP, we propose to use the newer QUIC protocol as the reliable transport for only
the handshake, resulting in QUIKE --- QUIC using IKEv2. As described in
Section~\ref{sec:bg_quic}, QUIC leaves the slot for its cryptographic handshake open
rather than mandating TLS~1.3, allowing us to fill it with IKEv2.

In contrast to TCP which is limited to providing transport features, QUIC requires
deeper integration as it brings mechanisms which were originally part of IKEv2. So,
QUIC subsumes the following IKEv2 mechanisms in QUIKE:
\begin{description}
	\item[Symmetric message protection.] Both the encryption and the integrity
		protection from IKEv2 are handled by QUIC's alternatives in QUIKE. This means that
		IKEv2 still negotiates the keys but instead of applying them itself, it derives
		per-level packet-protection secrets and hands those to QUIC\@.
	\item[Dead peer detection.] QUIC brings its own ping and idle timeout
		method, replacing IKEv2's dead peer detection.
	\item[Anti-\ac{DoS} cookie.] QUIC's retry obsoletes the cookie mechanism in
		IKEv2.
	\item[IKEv2 retransmission.] QUIC brings its own loss recovery.
	\item[IKEv2 fragmentation.] Fragmentation happens inside the transport handling of QUIC.
	\item[TCP encapsulation~\cite{rfc9329}.] QUIC provides a reliable transport layer,
		removing the need for TCP encapsulation.
\end{description}
In addition to the listed mechanisms, the MOBIKE extension~\cite{mobike} used for
IP-address roaming of IKEv2 peers can also be replaced by QUIC's native roaming
mechanism but we did not fully specify these interactions.

In QUIKE, all symmetric encryption is delegated from IKEv2 to QUIC, hence, we take
the full unencrypted IKEv2 packets as the plaintext for QUIC. The encryption keys
for QUIC are derived from seeds computed by IKEv2 during its normal key derivation step.
The derivation of these seeds for QUIC replaces the derivation of the keys for symmetric
encryption and integrity protection in IKEv2's key schedule. The
authentication keys and the IKE\_AUTH in IKEv2 are untouched by QUIKE as only the
symmetric encryption is delegated to QUIC. The encryption keys for QUIC's initial level
are computed exactly as in QUIC with TLS 1.3 with the only exception of changing a
static \ac{KDF} label from ``tls13 '' to ``ikev2 ''.
QUIC uses a TLS extension to transmit its transport parameters which is missing in
QUIKE\@. Thus, in QUIKE we add a new \emph{QUIC\_TRANSPORT\_PARAMS} notify message to
the initial exchange. That placement automatically includes it in the authenticated
transcript, ensuring its authenticity after IKE\_AUTH finished.

Analogous to IPsec over TCP~\cite{rfc9329} and the composite TCP
draft~\cite{hybrid_reliable}, there is no legacy discovery. The first packet is already
a QUIC initial packet and an unsupporting peer discards it without recovery. So
both peers must support QUIKE and activate it in their configuration. Further, QUIKE
inherits QUIC's requirement that the path carries 1200-byte datagrams, making it
unusable on paths below that floor.

Unlike an HTTP/3 deployment, QUIKE carries no payload traffic inside the QUIC
connection. After completing the handshake, we install the ESP \acp{SA} into the
kernel and the ESP traffic flows outside of it---just like in classic IPsec. Hence,
QUIC serves only as the transport for the handshake. We propose two different
approaches for the actual message flow, analogous to the two different TCP-based
drafts~(cf. Section~\ref{sec:proposals}): the two-round approach spends one
\ac{RTT} per key agreement, as the additional key exchanges do, while the composite
one carries both shares in a single key exchange payload, as the composite draft
does. Which of the two variants applies is negotiated implicitly, by selecting either
separate key exchange algorithms or a single, potentially composite identifier, as
with the TCP proposals.

\subsubsection{Two-Round QUIKE}%
\label{sec:quike_intermediate}
\begin{figure}
	\centering
	\providecommand{\imagedir}{./}
\input{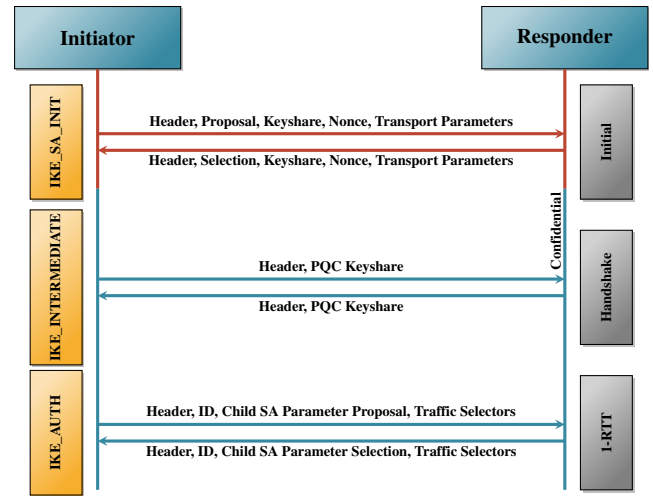}

\begin{tikzpicture}
	\def\width{450 pt}
	\def\rowsep{140}		
	\def\plainrun{115}		
	\def\encrun{290}		

	\node (init) [blue_box] {\huge\textbf{Initiator}};
	\node (resp) [blue_box] at ([xshift=\width] init)
		{\huge\textbf{Responder}};

	\draw[lifeline, draw=red] (init.south) -> ++(0,-\plainrun pt) coordinate(i1);
	\draw[lifeline] (i1) -> ++(0,-\encrun pt);
	\draw[lifeline, draw=red] (resp.south) -> ++(0,-\plainrun pt) coordinate(r1);
	\draw[lifeline] (r1) -> ++(0,-\encrun pt);
	\node[rotate=90, anchor=south east] at (r1)
		{\bfseries\Large Confidential};

	\coordinate (sa_init) at ([yshift=-70]init.south);
	\node[exchange] at ([xshift=-40]sa_init) {\bfseries\Large IKE\_SA\_INIT};
	\node[level] at ([xshift=\width+40]sa_init) {\bfseries\Large Initial};

	\draw[arrow, draw=red] ([yshift=8]sa_init) -> ++(\width,0)
		node[pos=0.5, above] {\bfseries\Large Header, Proposal,
		Keyshare, Nonce, Transport Parameters};
	\draw[arrow, stealth-, draw=red] ([yshift=-8]sa_init) -- ++(\width,0)
		node[pos=0.5, below] {\bfseries\Large Header, Selection,
		Keyshare, Nonce, Transport Parameters};

	\coordinate (intermediate) at ([yshift=-\rowsep] sa_init);
	\node[exchange] at ([xshift=-40]intermediate)
		{\bfseries\Large IKE\_INTERMEDIATE};
	\node[level] at ([xshift=\width+40]intermediate)
		{\bfseries\Large Handshake};

	\draw[arrow] ([yshift=8]intermediate) -> ++(\width,0)
		node[pos=0.5, above] {\bfseries\Large Header, \acs{PQC} Keyshare};
	\draw[arrow, stealth-] ([yshift=-8]intermediate) --
		++(\width,0) node[pos=0.5, below] {\bfseries\Large Header,
		\acs{PQC} Keyshare};

	\coordinate (ike_auth) at ([yshift=-\rowsep]intermediate);
	\node[exchange, minimum width=120] at ([xshift=-40]ike_auth)
		{\bfseries\Large IKE\_AUTH};
	\node[level] at ([xshift=\width+40]ike_auth) {\bfseries\Large 1-RTT};

	\draw[arrow] ([yshift=8]ike_auth) -> ++(\width,0)
		node[pos=0.5, above] {\bfseries\Large Header, ID, Child SA
		Parameter Proposal, Traffic Selectors};
	\draw[arrow, stealth-] ([yshift=-8]ike_auth) --
		++(\width,0) node[pos=0.5, below] {\bfseries\Large Header, ID,
		Child SA Parameter Selection, Traffic Selectors};

\end{tikzpicture}
	\caption{The message flow of two-round QUIKE, with the IKEv2 exchanges on
	the left and the corresponding QUIC levels on the right.}%
	\label{fig:quike}
\end{figure}
The two-round QUIKE resembles the structure shown in Figure~\ref{fig:intermediate}
but has exactly one intermediate exchange. This limits the amount of additional
\acp{KEM} to one. We assume that this protocol simplification's complexity reduction
justifies limiting the number of chained key agreements. Thus, each of the IKEv2
exchanges maps to one QUIC level, as portrayed by Figure~\ref{fig:quike}. In contrast
to IKEv2, already the first exchange is technically encrypted but it only provides
obfuscation, no confidentiality. The QUIC initial level carries the IKE\_SA\_INIT
including the small classical key exchange. It additionally contains QUIC's transport
parameters in the respective notify. Thereafter, the first symmetric keys are
derived and fed to QUIC's key ratchet. The intermediate exchange rides QUIC's
handshake level, containing the large \ac{PQC} \ac{KEM} data. The peers use the
exchanged secret to update the symmetric keys, providing QUIC with a new
per-direction pair of domain-separated, quantum-secure keys which protect the
subsequent messages. Finally, the 1-RTT level carries the IKE\_AUTH message,
authenticating the whole transcript. Further IKEv2 messages, like CREATE\_CHILD\_SA
exchanges, also run inside 1-RTT\@. The actual child \acp{SA}, on the other hand, go
into the kernel, allowing ESP traffic as in classic IPsec.

The alignment of QUIC's and IKEv2's levels/exchanges and key generations allows an
easier integration. Furthermore, QUIC's initial level faces several constraints to
impede amplification attacks~(cf. Section~\ref{sec:bg_quic}). Thus, the large
\ac{PQC} \ac{KEM} avoids those by riding the handshake level. The \acp{RTT} represent
those of the separate transports proposal~\cite{reliable_transport} but without the
preceding TCP handshake.

\subsubsection{Composite QUIKE}%
\begin{figure}
	\centering
	\providecommand{\imagedir}{./}
\input{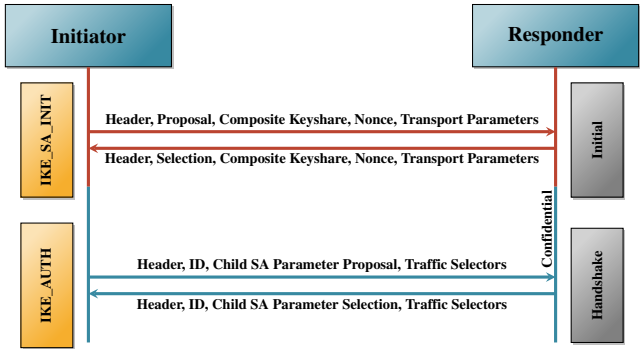}

\begin{tikzpicture}
	\def\width{450 pt}
	\def\rowsep{140}		
	\def\plainrun{115}		
	\def\encrun{150}		

	\node (init) [blue_box] {\huge\textbf{Initiator}};
	\node (resp) [blue_box] at ([xshift=\width] init)
		{\huge\textbf{Responder}};

	\draw[lifeline, draw=red] (init.south) -> ++(0,-\plainrun pt) coordinate(i1);
	\draw[lifeline] (i1) -> ++(0,-\encrun pt);
	\draw[lifeline, draw=red] (resp.south) -> ++(0,-\plainrun pt) coordinate(r1);
	\draw[lifeline] (r1) -> ++(0,-\encrun pt);
	\node[rotate=90, anchor=south east] at (r1)
		{\bfseries\Large Confidential};

	\coordinate (sa_init) at ([yshift=-70]init.south);
	\node[exchange] at ([xshift=-40]sa_init) {\bfseries\Large IKE\_SA\_INIT};
	\node[level] at ([xshift=\width+40]sa_init) {\bfseries\Large Initial};

	\draw[arrow, draw=red] ([yshift=8]sa_init) -> ++(\width,0)
		node[pos=0.5, above] {\bfseries\Large Header, Proposal,
		Composite Keyshare, Nonce, Transport Parameters};
	\draw[arrow, stealth-, draw=red] ([yshift=-8]sa_init) -- ++(\width,0)
		node[pos=0.5, below] {\bfseries\Large Header, Selection,
		Composite Keyshare, Nonce, Transport Parameters};

	\coordinate (ike_auth) at ([yshift=-\rowsep]sa_init);
	\node[exchange, minimum width=120] at ([xshift=-40]ike_auth)
		{\bfseries\Large IKE\_AUTH};
	\node[level] at ([xshift=\width+40]ike_auth) {\bfseries\Large Handshake};

	\draw[arrow] ([yshift=8]ike_auth) -> ++(\width,0)
		node[pos=0.5, above] {\bfseries\Large Header, ID, Child SA
		Parameter Proposal, Traffic Selectors};
	\draw[arrow, stealth-] ([yshift=-8]ike_auth) --
		++(\width,0) node[pos=0.5, below] {\bfseries\Large Header, ID,
		Child SA Parameter Selection, Traffic Selectors};

\end{tikzpicture}
	\caption{The message flow of composite QUIKE\@. The composite key exchange contains
		both key shares, so IKE\_AUTH moves to the handshake level, completing it before
	1-RTT.}%
	\label{fig:quike_composite}
\end{figure}

\label{sec:quike_composite}
Without the size constraint in IKE\_SA\_INIT (cf. Section~\ref{sec:bg_ikev2}),
removing the intermediate exchange and putting the \ac{PQC} \ac{KEM} into the first
exchange within a composite construct saves one additional \ac{RTT}.
Figure~\ref{fig:quike_composite} shows the resulting structure which resembles the
one of classic IKEv2 shown in Figure~\ref{fig:ikev2} and the one of classic QUIC in
conjunction with TLS~\cite{quic_tls}. IKE\_AUTH rides QUIC's handshake level. As the
\ac{PQC} \ac{KEM} already finished, it is protected with quantum-secure keys.
Despite including no message on the 1-RTT level during the handshake, 1-RTT is still necessary for
later exchanges, like creating new child \acp{SA}. From the trade-off it resembles the one of
the composite TCP proposal~\cite{hybrid_reliable}: We spare an \ac{RTT} for a larger
initial exchange. However, QUIC's constraints regarding its first messages (cf.
Section~\ref{sec:bg_quic}) may reduce the performance gain of this approach. Under
load a responder might then be even more likely to demand a retry, re-introducing the
saved \ac{RTT}. Hence, the performance gain could be more visible with smaller
\acp{KEM} where the \acp{RTT} still cost most of the time and could erode with
large \acp{KEM} where the initial level's constraints are more expensive.

\section{Evaluation}%
\label{sec:evaluation}
To answer \emph{\ref{rq:performance}} and \emph{\ref{rq:implications}}, we evaluate the different approaches within the
test environment described in Section~\ref{sec:measurement_setup}. Furthermore, we collect
findings from the implementation regarding its overall
complexity, the gap to the current codebase, and missing considerations within some
early drafts necessary for a subsequent holistic discussion. In this section, we
first describe the implementations, respective design choices, and practical
challenges in detail, before moving on to the measurement results and their
implications.

\subsection{Implementation}%
\label{sec:implementation}
For our implementation we used the
\emph{openiked-portable}\footnote{\url{https://github.com/openiked/openiked-portable}~[Accessed:
28.09.2026]}, which is a port of the \emph{openiked} of the \emph{OpenBSD}
project\footnote{\url{https://www.openbsd.org/}~[Accessed: 28.09.2026]}
to other operating systems, such as
Linux\footnote{\url{https://www.linux.org/}~[Accessed: 28.09.2026]}.
We implemented all drafts on top of the commit 4a71216cf3c6372a13451574d0ceb85c892e7376,
which at the time of starting this work was the most recent commit.
We used one branch per proposed mechanism but shared the commits for
shared prerequisites, keeping the implementations of the different approaches as
close and comparable as possible.
One prerequisite for all branches is the encoding of Classic McEliece's large public keys
which exceed the maximum length addressable by an IKEv2 payload header (cf.
Section~\ref{sec:prereq}). Of the three current proposals, we implemented the 32-bit
payload length draft using one reserved bit~\cite{big_payload}. As this approach is not yet
standardized, it does not yet have a status type identifier. We chose one from the
private-use range to allow negotiating it with supporting peers.

The first branch is the already standardized \ac{ADDKE}~\cite{rfc9370} together
with the intermediate
exchange~\cite{rfc9242} which are also the basis for all later approaches
except for the composite ones. Its standardization is already completed which gives a
well-defined specification, easing implementation. Additionally, we added the
codepoints for the \acp{KEM} from the private-use-range. However, in line with
previous analysis~\cite{real_ipsec}, signing a fragmented intermediate exchange
requires complex code changes. The reason for this is the requirement that the
receiver must sign the exchange as if it was not fragmented which means it needs to
reconstruct the message without any fragmentation artifacts instead of immediately
processing it, contradicting existing implementation flows.

\begin{table}[!tb]
\caption{Open design choices of both acknowledgment
  proposals~\cite{fragment_ack,selective_retransmission}.}%
\label{tab:decisions}
\centering
\begin{tabularx}{\columnwidth}{@{}p{0.16\columnwidth}XX@{}}
\toprule
\textbf{Decision} & \textbf{Our choice} & \textbf{Consequence} \\
\midrule
Trigger \mbox{(status-ACK)}
	& A quiet period of $500$\,ms (above \ac{RTT}), then every $500$\,ms
	& Below the \ac{RTT} it triggers prematurely, far above it is slow  \\
\addlinespace
Reordering \mbox{(status-ACK)}
	& The same quiet period after reordering
	& A late straggler causes a premature trigger \\
\addlinespace
Trigger \mbox{(notify-ACK)}
	& After receiving a retransmitted fragment
	& Waits for a retransmission; fragments still in flight count as missing \\
\addlinespace
Regular \mbox{retransmit}
	& ACKs restart its timer
	& ACKs without progress defer give-up until an absolute deadline \\
\addlinespace
Key retention
	& Retain the previous IKE \ac{SA} keys
	& Required to decrypt ACKs \\
\bottomrule
\end{tabularx}
\end{table}

The next branches are the two retransmission
optimizations~\cite{fragment_ack,selective_retransmission}. Both drafts leave design
decisions to the implementer, especially regarding the exact definition of when to send
their form of acknowledgment. Table~\ref{tab:decisions} shows our choices for those
gaps. For status-ACK, a receiver sends a status message once no new fragment of an
incomplete message arrived for a quiet period of $500$\,ms, just above the largest
\ac{RTT} of our grid, and repeats it every $500$\,ms while fragments are missing. On a
path whose \ac{RTT} exceeds that period, it sends acknowledgments before a response
can arrive, causing redundant retransmissions. For notify-ACK, we follow the draft's
conservative suggestion and only acknowledge after receiving a retransmitted fragment.
Neither draft specifies how acknowledgments interact with the classical
retransmission mechanism following an exponential backoff timer. In our
implementation, an acknowledgment restarts that timer and for status-ACK lets the sender drop
acknowledged fragments from its retransmission queue.

Implementing both drafts further revealed two gaps that neither addresses. First,
\emph{previous-key retention}: in \ac{ADDKE}, the initiator can only derive the next
IKE \ac{SA} keys after receiving the complete response~\cite{rfc9242,rfc9370}, so its
acknowledgments for missing response fragments are still protected with the previous
keys, which the responder thus has to retain beyond its own key update. Second,
neither draft defines how acknowledgments affect the retransmission limit. If an
acknowledgment restarts the retransmission timer, acknowledgments without progress
keep the initiator from giving up until an absolute handshake deadline, which
RFC~7296 does not mandate~\cite{ikev2}. If acknowledgments are sent periodically and
some fragments cannot reach the peer at all, this can even happen with honest peers.

The \ac{FEC} approach utilized RaptorQ fountain codes, as described in RFC~6330~\cite{rfc6330}.
For implementation, the open-source library
\emph{nanorq}\footnote{\url{https://github.com/sleepybishop/nanorq} [Accessed: 02.10.2026]}
was used which allows for easy en- and decoding of IKE\_INTERMEDIATE messages.
Plain fragments act as source symbols to the encoder, producing many encoding symbols,
which are sent over the wire instead of plain message fragments. The encoder outputs a total of $s+o+e$ symbols, where $s$ is the number of source symbols, $o$ is the fixed number of overhead packets needed due to the probabilistic design of fountain codes, and $e$ is the number of extra packets always taken as a fraction of $s$, adding redundant symbols acting as the resilience cushion.
Decoding starts as soon as at least $s+o$ encoding symbols arrive,
to maximize the likelihood of success.
The test environment used $5$ overhead and $10\%$ extra packets.

It is important to note that the redundancy only protects the message itself.
If the first fragment is lost in transmission, the exchange cannot proceed as it
contains information about the inner IKEv2 payload type. This inner payload type is
required to tell the IKEv2 message parser the payload type of the first following
payload. Each payload carries this information for the following one. This means a
successful decoding of the actual payload stream is irrelevant for further processing
if the type of payload cannot be determined. Adding payload type information to every
fragment would violate RFC~7383~\cite{ike_frag}.

The two TCP-based proposals rely on the kernel's TCP implementation which provides
a robust transport of the messages. Hence, with TCP we never use IKEv2
fragmentation. TCP provides reliability but requires the application to manage its
connection state and control its flows: the connection management, the mapping
to the respective \acp{SA}, a handler for partial writes caused by full TCP write
buffers, an accumulator reassembling the byte stream into IKEv2 messages, a TCP
accept handler, and a watchdog for connections without an \ac{SA} are left to the
IKEv2 implementation. Additionally, we added a necessary stream prefix and additional
message length field of RFC~9329~\cite{rfc9329}. The composite approach further needs
new identifiers for the composite key exchange, including its implementation and
combiner. The other one requires the \ac{ADDKE} logic including intermediate
exchanges but without fragmentation, easing their implementation. Our implementations
do not include negotiation (like proposed by the separate transports
draft~\cite{reliable_transport}) but assume all peers begin their handshake over
TCP\@.

To implement the QUIKE proposals, we first introduced the QUIC library
\emph{ngtcp2}~$1.24.0$\footnote{\url{https://github.com/ngtcp2/ngtcp2} [Accessed: 29.09.2026]}
which has a comparatively loose coupling of QUIC and TLS, easing the IKEv2
integration. However, to make it compatible with our tests, we had to adjust four of
its limits. The adjustments include raising the maximum size of the initial
(composite QUIKE) and handshake level (two-round QUIKE), as well as the maximum
reordering gap, all necessary to carry a Classic McEliece public key. None of these is
mandated by the standard but chosen by the library, avoiding deviations from QUIC's
specification~\cite{quic}. The fourth change is motivated by FrodoKEM and enlarges
the initial congestion window to $30$ datagrams. This allows the peers to send exactly one full
FrodoKEM message in one flight, without waiting for the peer's acknowledgment of the
first packets, saving one unnecessary \ac{RTT}. It deviates from a specification's
SHOULD~\cite{quic_recovery} which proposes ten datagrams, based on
RFC~6928~\cite{rfc6928}. RFC~6928 recommends this window size based on measurements with
applications aggressively opening concurrent connections causing congestion. The
\ac{VPN} usually initiates exactly one and still has more congestion control than the
UDP-based approaches, legitimizing this change. In addition to the adjustments, we
also fixed one ngtcp2 bug where lost ping messages could congest a send-queue.
Merging the fix upstream is future work.

For the QUIC integration we took the patched library and added an interface module.
It gets the plain text messages and domain separated keys from IKEv2 to encapsulate
them in the QUIC messages and transmit those reliably, or, on the receive path, hand
IKEv2 the decrypted IKEv2 messages. Both the QUIC implementation and the interface
module could theoretically live inside a separate process which then only sees its
own keys but not the other domain separated ones for authentication or the ESP
\acp{SA}. This would harden the implementation as it shields
\emph{openiked} from the large amount of new code. However, in our testing
implementation we kept all modules in the \emph{openiked} process, deferring that separation to future work.

\subsection{Results}%
\label{sec:results}
The tests for answering \emph{\ref{rq:performance}} span, for the matrix of different mechanisms,
\acp{KEM}, and network parameters, $1296$ different measured configurations. For each
configuration, we measured between $80$ and $200$ handshakes, depending on their
decisiveness regarding completion rate and duration. Even though we set the minimum number of runs in our testbed to 40, no configuration was decisive enough after fewer than 80 runs.
Failures are rare and confined to the two highest
loss levels: $101$ of $126\,500$ handshakes failed after timing out or reaching a
retransmission limit, none below $5\%$ packet
loss. $76$ of them happened in Classic McEliece, $24$ in FrodoKEM, and $1$ in ML-KEM.
Our replication package contains the exact data for all runs~\cite{repl_package}.

\begin{figure}[!t]
	\centering
	\includegraphics[width=\columnwidth]{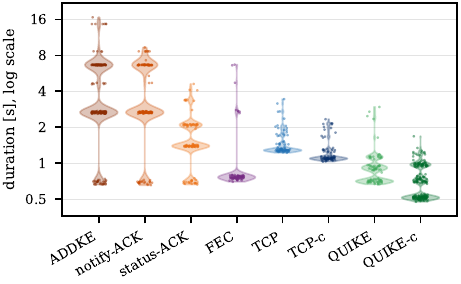}
	\caption{Logarithmically ordered completion times of the FrodoKEM measurements over
		a $10$\,Mbit/s link at $100$\,ms one-way delay and $2\%$ packet loss. Each
		implementation has between $80$ and $200$ runs, and \emph{-c} marks the
		composite variants. Violin plots for adjacent
		configurations are in Appendix~\ref{app:violin}.}%
	\label{fig:res_distributions}
\end{figure}
Before calculating further statistics, we need to inspect each approach's data
structure, as it varies significantly.
Figure~\ref{fig:res_distributions} shows the distribution of handshake
times on a logarithmic scale for one medium-network, medium-loss FrodoKEM
configuration. Other configurations and \acp{KEM} exhibit a similar structure, as
shown in the figures of Appendix~\ref{app:violin}. The
first clear observation regarding the data structure is that the results land in
discrete plateaus, with the
transport-based approaches having significantly more stragglers between plateaus and
the plateaus being closer together. The reason for that lies in the retransmission
logic. Basic RFC~7383~\cite{ike_frag} sends all fragments of a message and, receiving
no answer, retransmits all of them after a blind timer that doubles with every
attempt. A run's duration is therefore roughly its loss-free duration plus a sum of those
intervals, and which sum it lands on depends on two independent counts: how often a
single message had to be retried, and how many different messages were hit at all.
Losing several fragments of one flight still costs a single retry of $2$ seconds,
whereas losing one fragment in each of two messages costs a retry for each, and a
retransmission that is hit again costs $2+4$ seconds. The same logic applies to
\ac{FEC}. There, the first message is larger and slower but its redundancy prevents
retransmissions. Hence, its plateau is a bit later but it almost never needs
retransmission. The other two datagram approaches follow a similar logic until their
own acknowledgment-based logic takes over, with the status-ACK doing so more
aggressively as it already sends its acknowledgment after not receiving any more fragments for 500\,ms (cf. Table~\ref{tab:decisions}) while the notify-ACK sends it only after getting the first retransmission. The transport-based approaches, on the other hand, react immediately to
loss, hence the clusters' distance comes from the \ac{RTT}. More loss causes more
\acp{RTT} until completion. For FrodoKEM, we can further see that the adjusted
initial window size of QUIKE (cf. Section~\ref{sec:implementation}) saves another
\ac{RTT} in comparison to TCP\@.

This distribution discourages the use of percentile-based metrics, like the median,
for the comparison. The median and its confidence are still well-suited to determine
clear configurations, usually with low loss. Yet in a comparison, such a value could
lie just at the bottom of a cluster and just at the top for another approach, making
them falsely look equal. At the same time, it could be apart by a factor of two
between two almost equally fast approaches. Instead, in the following we use the mean as the stronger
indicator for an approach's performance. There, failed runs could potentially distort
the results as we exclude them from the mean, biasing for quicker run times in
low-success configurations. However, the low overall failure rate confines this issue
to only a few data points. Further, the mean duration of low-success configurations is also
especially high. Thus, this does not affect any deductions.
Appendix~\ref{app:stacked} contains all means and their respective cell's failure
rate for deeper scrutiny. From the different approaches' structure and their means,
we can make the following deductions:
\begin{description}
	\item[Datagram \& Transport.] The approaches split into the two classes of datagram-
		and transport-based ones.
	\item[Latency \& Loss.] A higher latency mostly affects the loss-recovery of the
		transport-based approaches, while the others' mechanisms depend mainly on their
		own timers.
	\item[Latency \& Bandwidth.] Latency affects smaller \acp{KEM} and bandwidth limits
		the larger ones. Varying the bandwidth changes the mean by a median factor of
		$1.11$ for ML-KEM and $7.39$ for Classic McEliece, varying the latency by
		$3.29$ and $1.73$ respectively.
	\item[Loss \& Congestion.] Datagram-based approaches do not control any congestion.
		Transport-based ones interpret loss as congestion, sometimes unnecessarily
		limiting their own bandwidth.
	\item[Loss Position.] Loss spread over multiple IKEv2 messages costs more
		than the same amount confined to one. For the transport-based ones, early
		congestion events have more subsequent fragments to hamper than later ones.
	\item[\ac{KEM} Size.] Across configurations, the \ac{KEM} size is a
		stronger factor than the approach.
\end{description}

\subsubsection{ML-KEM and FrodoKEM}%
\label{sec:res_common}
\begin{figure*}[t]
	\centering
	\includegraphics[width=\textwidth]{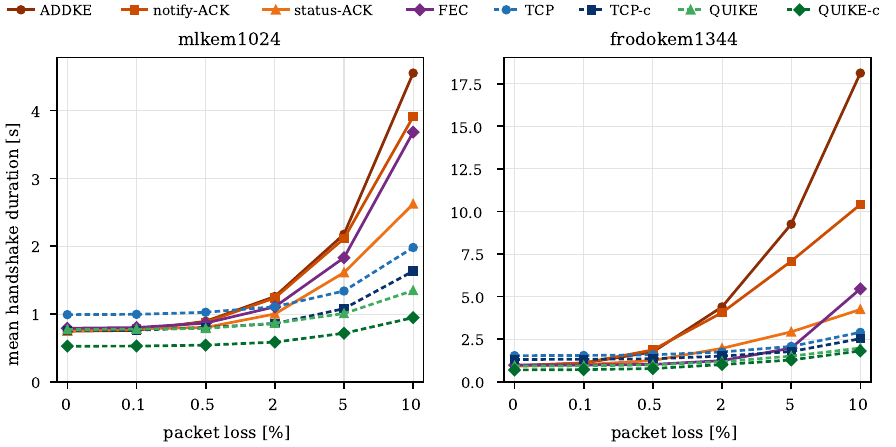}
	\caption{Mean handshake duration across all bandwidths and latencies against packet loss for ML-KEM and
	FrodoKEM\@. Solid lines depict datagram-based approaches, dashed lines
	transport-based ones. Lines connect the measured loss levels and are not
	interpolation.}%
	\label{fig:res_common}
\end{figure*}

All results for ML-KEM and FrodoKEM follow a similar structure, depicted in 
Figure~\ref{fig:res_common}. The data shown in Figure~\ref{fig:res_common} is computed in
two steps: first, we compute the mean handshake duration for each configuration (cf. Table~\ref{tab:grid}), and 
second, for each combination of \ac{KEM} and network loss we compute the mean of the mean
handshake durations across all latencies and bandwidths.
Nesting means gives each of the pooled means equal weight despite differing amounts
of runs. Still, the different network configurations produce varying durations, which
Appendix~\ref{app:stacked} shows in Figure~\ref{fig:app_grid}.

The data shows that with lower loss, the amount of necessary \acp{RTT} dominates the
handshake duration. All datagram-based approaches fare equally in that scenario with
also the additional overhead of FEC having no significant impact. TCP keeps the additional
\ac{RTT} for the IKE\_INTERMEDIATE and adds the TCP handshake, slowing it down. For FrodoKEM,
both TCP variants pay another \ac{RTT} as it is limited by the kernel's initial
congestion window whereas QUIKE is optimized to fit exactly one FrodoKEM message,
setting those two families apart.
TCP-composite eliminates the intermediate \ac{RTT}. QUIKE still has it but
QUIKE-composite spares all of them, making it the fastest approach across
configurations.

With increasing loss, the transport-based approaches overall cope significantly better than almost all datagram-based ones. The only
exception is FEC for FrodoKEM, where its redundancy absorbs most loss until $5\%$. Only
afterwards its slow retransmission logic bites. This is due to FEC's redundancy being a fixed fraction
of the message---10\% in our testbed---so it only covers the losses reliably as long as the margin between the experienced loss and the added redundancy is big enough.
With ML-KEM in contrast, there is only a single redundant packet because of the smaller message sizes, which a second loss already defeats, making its performance
significantly worse there. The remaining
three datagram approaches behave consistently under higher loss. The basic ADDKE one
struggles most, notify-ACK's effect only shows significantly for FrodoKEM under high
loss, and status-ACK is overall closest to the transport-based ones.

\subsubsection{Classic McEliece}%
\label{sec:res_mceliece}
\begin{figure*}[pbt]
	\centering
	\includegraphics[width=\textwidth]{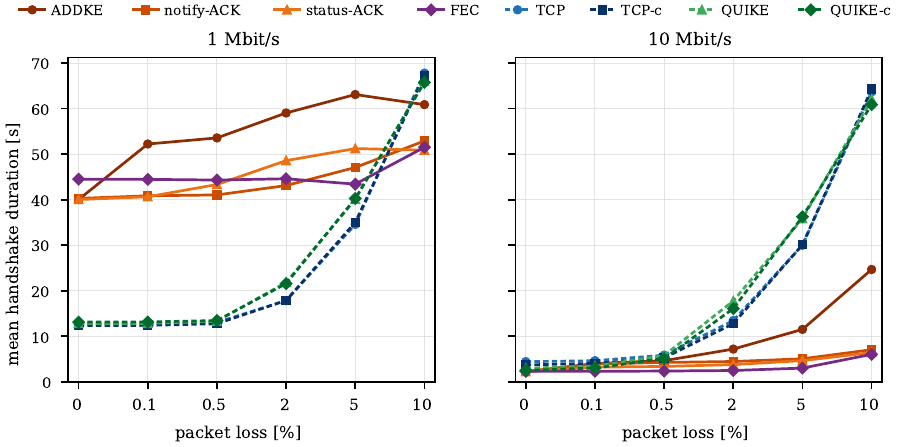}
	\caption{Mean of the per-latency means of the handshake durations at a given packet
		loss for Classic McEliece at $1$ and $10$\,Mbit/s. The exact crossing point of
		the transport- and the datagram-based approaches for $1$\,Mbit/s depends on the
		latency. Appendix~\ref{app:stacked} shows the complete grid.}%
	\label{fig:res_mceliece}
\end{figure*}

The Classic McEliece results differ from the previous ones as network congestion or
mechanisms to avoid it influence them. This impact changes significantly between the
$1$ and $10$\,Mbit/s measurements. Figure~\ref{fig:res_mceliece} shows the mean over
the means of the different latencies for each loss configuration and the two
bandwidths. For the smaller bandwidth, no datagram-based approach manages to be
significantly faster than $40$\,s. At this bandwidth, serializing one Classic McEliece
public key alone takes about $11$ seconds. Their blind, exponential retransmission
timer therefore expires twice while the first copy still drains, sending the flight
three times over and congesting the path. The subsequent authentication message needs to wait until the link is free
again. FEC takes even longer as with every retransmission it also resends its
redundancy packets. The loss, on the other hand, has less effect, as the biggest
message is retransmitted several times anyway, reducing the probability of further
retransmissions being necessary. FEC's redundancy, again, absorbs the loss up to
$5\%$, whereas the acknowledgment-based proposals degrade throughout while still
staying significantly ahead of \ac{ADDKE}\@. Here FEC's fixed fraction works in its
favor: roughly 3000 fragments carry close to 300 redundant symbols
against an expected 150 losses at $5\%$, so it only breaks down at
$10\%$ once the expected losses are roughly the same as the number of redundant symbols. As soon as the bandwidth
suffices to transmit every message before any blind timer triggers, the congestion
events vanish and the datagram-based approaches' duration collapses.

The transport approaches avoid congestion entirely as they have integrated congestion
control mechanisms. Yet, at low bandwidth and no loss, their initial congestion
control window is still small, limiting the first messages, costing \acp{RTT}. A
lower latency shortens the \acp{RTT}, reducing the initial window's impact on
duration. They, however, interpret higher loss as congestion and, in response, limit
their sending speed significantly. In detail, they reduce the window size, which
means they need to wait more often for acknowledgments. High latency increases these
waiting times, further exacerbating the problem. This follows the Mathis model, which
predicts a throughput proportional to $\mathit{MSS}/(\mathit{RTT}\sqrt{p})$ for a
\ac{MSS} and a loss probability $p$~\cite{mathis1997}, so
the durations at high loss and latency are congestion control working as specified.
As this slowdown scales with the RTT, the durations of the transport-based approaches
diverge strongly across latencies at high loss whereas the datagram-based ones stay
almost constant. The pooled mean therefore represents none of the latencies well
and, at $1$\,Mbit/s, hides that the transport-based approaches are the fastest at low
latency. The latency-specific durations cross the datagram-based approaches at
different points and are shown in Appendix~\ref{app:stacked}. Moreover, the means of
the high-loss, high-latency TCP configurations are biased downward, as these
configurations produced the highest share of failed handshakes, $62$ of Classic
McEliece's $76$, which are excluded from the mean.
In contrast to the actual congestion events of the datagram-based approaches, the
rate-limiting by congestion control is purely loss-driven and, therefore, independent
of the actual bandwidth.

\begin{figure}[pbt]
	\centering
	\includegraphics[width=\columnwidth]{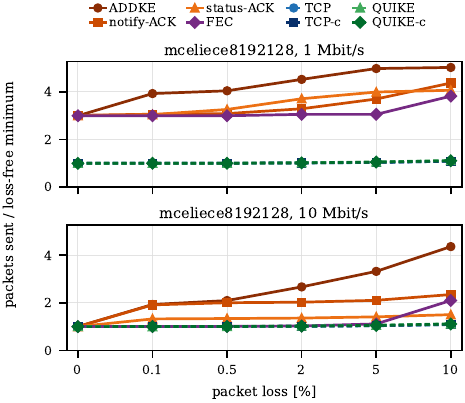}
	\caption{Packets the initiator sends for Classic McEliece, each approach relative
		to itself on an uncongested loss-free path. Appendix~\ref{app:efficiency} shows
	  the same information for ML-KEM and FrodoKEM.}%
	\label{fig:res_overhead}
\end{figure}
The amount of transmitted packets also shows the congestion.
Figure~\ref{fig:res_overhead} shows the ratio of the packets actually sent by the
initiator divided by the minimum number of packets for that approach, without loss
and without congestion. So it shows how the environment changes the packet count, not
fixed overhead, like FEC overhead or static padding for transport frames (cf.
Section~\ref{sec:implementation}). With Classic McEliece and $1$\,Mbit/s, all
datagram-based approaches retransmit at least two times, even without loss, making
them start at factor three already. As soon as the link is broad enough to avoid
congestion, their structure reflects the one of the duration; hence, a slower
approach is also less efficient. This also resembles the efficiency of the other two
\acp{KEM}. So, without congestion and across \acp{KEM}, basic \ac{ADDKE}'s primitive
logic copes the worst with difficult conditions. Notify-ACK mostly follows second; a
faster link even weakens its efficiency as its more cautious notify mechanism is then
too slow to reduce the amount of sent packets. FEC's redundancy absorbs loss until a
predictable threshold, dependent on its configuration; and status-ACK's aggressive
mechanism and queue-pruning are most successful at preventing redundant
retransmissions.

While the transport approaches for McEliece slow down significantly under heavy loss,
their efficiency remains almost unchanged. This observation is stable across
bandwidths and latencies. Congestion control prevents overloading a link but is
sometimes too cautious after observing losses, costing \acp{RTT} but preventing
redundant transmissions. Also this observation is in line with the other \acp{KEM}
(cf. Appendix~\ref{app:efficiency}), where increases in duration are also caused by
the initiator waiting for acknowledgments, never redundant retransmissions.

\section{Discussion}%
\label{sec:discussion}
In this section, we build on the results of Section~\ref{sec:evaluation}, put them
into perspective and then provide answers to \emph{\ref{rq:implications}}. To do so, we first lay out
how the approaches depend on each other and what implementing them costs. Then, we
regard the state of Classic McEliece integration. Further, we consider the situation for the
other \acp{KEM} and what it means for research, standardization, and especially
implementation. Finally, we outline important trajectories for future work.

The results in Section~\ref{sec:results} show that for each \ac{KEM}'s size regime,
the approaches significantly differ in performance and resilience. However, the main
factor remains the \ac{KEM} size itself. Thus, analyzing the risk and determining
which \ac{KEM} size should be supported is the first step. Afterwards, the path of
standardization and implementation needs to be decided, optimized for exactly this
choice of \acp{KEM}.

\subsection{Approach Structure and Implementation Cost}%
\label{sec:disc_structure}
\begin{figure}[!t]
	\centering
	\resizebox{\columnwidth}{!}{\providecommand{\imagedir}{./}
\input{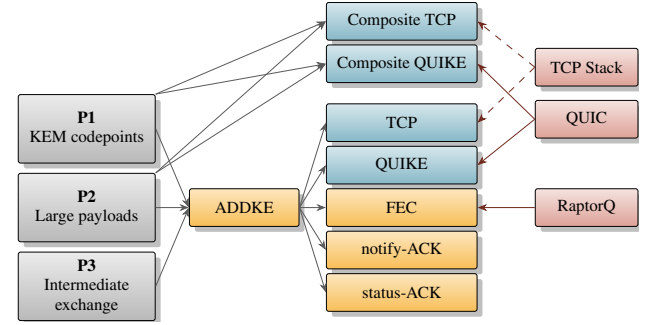}

\begin{tikzpicture}[
	font=\scriptsize,
	gbox/.style={draw=black, line width=0.4, align=center, inner sep=2pt,
		rounded corners=1pt, angle, drop shadow},
	pre/.style={gbox, minimum width=2.0cm, minimum height=1.0cm,
		top color=gray!15, bottom color=gray!55},
	app/.style={gbox, minimum width=2.2cm, minimum height=0.55cm,
		top color=blue!20, bottom color=blue!60},
	dgram/.style={app, top color=green!30, bottom color=green!80},
	hinge/.style={dgram, minimum width=1.6cm},
	ext/.style={gbox, minimum width=1.5cm, minimum height=0.55cm,
		top color=red!15, bottom color=red!50},
	req/.style={draw=black!65, line width=0.4, -{Stealth[length=4pt]}},
	extreq/.style={req, draw=red!60!black},
	kernreq/.style={extreq, dashed},
]
	\node[pre]   (p1)    at (1.0,  0.43) {\textbf{P1}\\KEM codepoints};
	\node[pre]   (p2)    at (1.0, -0.72) {\textbf{P2}\\Large payloads};
	\node[pre]   (p3)    at (1.0, -1.87) {\textbf{P3}\\Intermediate\\exchange};

	\node[hinge] (addke) at (3.3, -0.72) {ADDKE};

	\node[app]   (ctcp)  at (5.6,  2.00) {Composite TCP};
	\node[app]   (cquik) at (5.6,  1.38) {Composite QUIKE};
	\node[app]   (septr) at (5.6,  0.52) {TCP};
	\node[app]   (quik)  at (5.6, -0.10) {QUIKE};
	\node[dgram] (fec)   at (5.6, -0.72) {FEC};
	\node[dgram] (nack)  at (5.6, -1.34) {notify-ACK};
	\node[dgram] (sack)  at (5.6, -1.96) {status-ACK};

	\node[ext]   (tcp)   at (8.3,  1.33) {TCP Stack};
	\node[ext]   (quic)  at (8.3,  0.57) {QUIC};
	\node[ext]   (rq)    at (8.3, -0.72) {RaptorQ};

	\draw[req] (p1.east) -- (addke.west);
	\draw[req] (p2.east) -- (addke.west);
	\draw[req] (p3.east) -- (addke.west);
	\draw[req] (addke.east) -- (septr.west);
	\draw[req] (addke.east) -- (quik.west);
	\draw[req] (addke.east) -- (fec.west);
	\draw[req] (addke.east) -- (nack.west);
	\draw[req] (addke.east) -- (sack.west);

	\draw[req] (p1.north east) -- (ctcp.west);
	\draw[req] (p2.north east) -- (ctcp.west);
	\draw[req] (p1.north east) -- (cquik.west);
	\draw[req] (p2.north east) -- (cquik.west);

	\draw[kernreq] (tcp.west) -- (ctcp.east);
	\draw[kernreq] (tcp.west) -- (septr.east);
	\draw[extreq] (quic.west) -- (cquik.east);
	\draw[extreq] (quic.west) -- (quik.east);
	\draw[extreq] (rq.west)   -- (fec.east);
\end{tikzpicture}}
	\caption{The dependencies of the different approaches. \acs{ADDKE} is both an
		approach and a prerequisite for others. Yellow approaches are datagram-based,
		blue ones transport-based. External dependencies are shown in red, with a dashed
	  arrow marking kernel-based ones.}%
	\label{fig:depgraph}
\end{figure}

Figure~\ref{fig:depgraph} shows the dependencies between the approaches. Every
approach requires P1 and P2, yet P2 is only necessary for Classic McEliece-sized
keys. P3 is the foundation for \ac{ADDKE} which, in turn, is a prerequisite for every non-composite
approach. This is especially important as \ac{ADDKE} and P3 together are, considering
lines of code, the biggest change with $2545$ or $3122$ lines considering all
prerequisites, with most located within existing
modules. The second largest change is composite QUIKE with $1700$ (alone) or $2277$
(including P1/2) lines, and the largest share put into separate modules. TCP is
smaller but its changes are again more intertwined with existing software. Every non-composite
approach then has a smaller footprint on its own but that is on top of \ac{ADDKE}.
Those numbers give an intuition about the implementations' intrusiveness but do not
necessarily reflect their complexity. E.g., signing fragmented intermediate exchanges
was an especially complex task without needing much code. They also do not consider conceptual challenges, like unfinished
standardization, previous-key retention, or delegating confidentiality to a different
protocol.
Replacing existing parts with external dependencies, on the one hand, leads to large
and complex changes but, on the other hand, might make the final design simpler overall.
Further, the external dependencies, like the whole QUIC stack, can introduce large
pieces of software with their own security flaws, potentially degrading the overall
security. Hence, integration requires thorough evaluation. TCP is less critical
as the kernel provides it already.

\subsection{Classic McEliece}%
\label{sec:disc_mceliece}
While Classic McEliece is currently not considered as a widespread solution for ephemeral
\acp{KEM} (cf. Section~\ref{sec:bg_kems}), it still demonstrates the maturity of an
approach, should \ac{KEM} sizes overall increase into that size regime. Our results
shown in Section~\ref{sec:results} make clear that, as soon as the network conditions
worsen, every approach either took over a minute for one handshake on average, or
congested the network, practically denying service to any other device using that
link. We consider none of these operationally acceptable. Hence, right now such large
\acp{KEM} are only suitable for specific use cases where the network conditions are
stable, e.g., on a dedicated, high-performance layer two network~\cite{pqc_macsec}. Apart from that,
further research needs to find ways to, either, prevent the congestion for the
datagram-based approaches, or soften the effect of the transport-based approaches'
slow start and congestion control, both without introducing new caveats.

\subsection{Two Paths for ML-KEM and FrodoKEM}%
\label{sec:disc_paths}
For the other two \acp{KEM} there are two fundamental paths for implementations to
follow:
\begin{description}
	\item [Established Path.] This represents the compatible option. It builds on the
		already standardized \ac{ADDKE}, allows step-wise optimization, and upholds
		backwards compatibility but pays in capability and a more patchwork design.
	\item [New Transport.] This delegates reliability, congestion control, and, with
		QUIKE, framing and packet protection. Its best variant, composite QUIKE, is
		always at least as fast as any datagram-based approach, but it introduces new
		dependencies and currently lacks standardization.
\end{description}
Taking the established path adds heavy machinery that is usually part of transport
protocols. However, it builds on an already standardized base, reusing the same
foundation of classic IKEv2 and reducing friction by, e.g., allowing
backwards compatibility, keeping testing infrastructure, firewall rule sets, etc.
Further, it allows step-wise adoption: For ML-KEM, \ac{ADDKE}'s performance might still be
deemed acceptable. We show that further optimizations
significantly improve performance under more challenging network conditions. We deem
this helpful for ML-KEM and mandatory for FrodoKEM-sized keys. The optimizations
include:
\begin{description}
  \item [Status-ACK.] This approach adds an acknowledgment mechanism based on status
		messages. The implementation does not add significant hurdles to \ac{ADDKE} and
		the results show a strong performance increase compared to \ac{ADDKE} under
		worsening networking conditions. Additionally, its explicit receipt bitmap
		allows pruning retransmit queues, making retransmits more efficient and adding
		less redundant traffic. It is
		still in the early standardization phase and its security considerations
		address neither the mandatory previous-key retention nor the retransmission limit
		(cf. Section~\ref{sec:implementation}). The former means a responder still
		accepts acknowledgments protected only by the classical IKE\_SA\_INIT keys after
		the first \ac{PQC} key exchange, the latter lets a malicious peer delay the
		initiator's give-up.
	\item [Notify-ACK.] This approach falls into the same class as status-ACK but
		is generally slower and less efficient, and it shares both gaps. Additionally, its
		notifies carry no message counter, so an on-path attacker can replay them, each
		time making the initiator resend request fragments, which IKEv2 initiators
		otherwise only do on timeout. Also, there is an open issue with nonce
		reuse, breaking confidentiality and authenticity for that \ac{SA}. In conclusion,
		we see the status-ACK proposal as overall the better alternative for this kind of
		optimization approach.
	\item [\ac{FEC}.] Instead of optimizing retransmission, \ac{FEC} tries to prevent
		it in the first place. It does not provide the same benefit as status-ACK for
		small \acp{KEM} but mostly fares better for FrodoKEM\@. It has the simplest
		mechanic of all approaches (not considering preconditions) and fewer protocol
		interactions but an additional dependency. For a software project willing to accept the
		additional dependency, this is a relatively cheap optimization for \ac{ADDKE}
		with medium-sized \acp{KEM}.
\end{description}
An implementation choosing continuity over maximum performance chooses this
datagram-based path. It first adds \ac{ADDKE} and then, if necessary, optimizes it
with a finalized version of status-ACK or \ac{FEC} (for FrodoKEM or similar \acp{KEM}).
Further, both mechanisms are not mutually exclusive. So, even combining them is
plausible with the aim to both reduce the chance of retransmissions and make them
more efficient and effective where they are still necessary.

The transport path is the alternative way of moving forward. It critically changes
core design choices for IKEv2, instead of working around them. The result is an
overall leaner design with clearer delegations and better performance but at the cost
of continuity. The four variants split into composite or two-round and TCP or QUIKE.
When choosing between them, our results give a clear picture regarding the first
choice: Composite constitutes a significantly smaller change (cf.
Section~\ref{sec:disc_structure}) and performs better.

Choosing between (composite) TCP and QUIKE is less clear. In general, QUIKE resembles
the more extreme variant. It changes the protocol more fundamentally, delegates more
to the new transport, introduces a new external dependency, and, due to its novelty,
lacks standardization. That shows in the change size: its implementation is larger,
albeit more localized to its module which is the interface between IKEv2 and QUIC,
while TCP is smaller but more distributed throughout the codebase. Additionally,
QUIC's retry token and its cap on what a peer may send to an unvalidated address
represent the most advanced version of \mbox{pre-authenticated} \ac{DoS} protection.
Our implementation includes a bug fix of the introduced QUIC stack, highlighting the
maturity risk of such a dependency. However, the clear separation between QUIC and
IKEv2 theoretically allows to put both implementations into separate processes, where
the QUIC process only ever sees its own domain-separated keying material, not the
other keys used for authentication and protecting the application data. Hence, such
an implementation could limit the potential damage by a vulnerable QUIC stack.

In summary, there are two main pathways: Step-by-step working around design decisions
aimed towards small, classical \ac{DH} shares but keeping continuity regarding
implementations, test setups, firewall rule sets, and backwards compatibility with
existing setups; or, radically changing those design decisions to match the current
context but causing disruption. TCP, thereby, resembles the less radical of the two
variants. Nevertheless, we deem the advantages of the novel QUIKE worth the little
increase in disruption, compared to TCP\@. Implementers, however, need clear
specifications, which to date leaves \ac{ADDKE} as the only viable option, despite its
weaker performance.
So the IETF must either decide between them or standardize both pathways, allowing
the implementations their own choice.

\section{Threats to Validity}%
\label{sec:threats_to_validity}
In this section, we discuss potential threats to the validity of our study and how we
mitigated them. Its structure adheres to the categories of Ampatzoglou et
al.~\cite{Ampatzoglou20ThreatsToValidityGuidelines}: study selection validity, data
validity, and research validity.

Our systematic proposal search described in Section~\ref{sec:sys_review} focuses only on
IETF drafts present in the IETF corpus' snapshot.
We could miss pure academic work. Also the snowballing
only included RFCs and drafts, even excluding IETF documents with deviating draft
citation tags. Yet, we only found two IETF documents excluded by the snowballing and
they were already in the screened set. Also a manually checked sample of the
referenced academic literature did not yield any new proposals but foundational
work and analyses. We only had a single screener which we mitigated by applying
criteria fixed before screening and publishing the complete decision record.

The data validity depends on our implementations of the current Internet-Draft states and their tuning. Errors there could
limit specific drafts. We set timers and windows to values trying to optimize each
mechanism while keeping operational feasibility (e.g., a $500$\,ms timer for
acknowledgment for a $400$\,ms \ac{RTT}). Our test harness uses netem with
random, uncorrelated loss. The random loss has less statistical variance
which makes the results more stable and comparable but real loss often occurs in
bursts~\cite{loss_burst}. Also we record netem's
packet counter but not its byte counter. Hence, different packet sizes or
padding fluctuations could distort the efficiency results. In response, we only
compared each arm against its normalized best case, mitigating the effect of
differing packet sizes. Different padding might still affect the results.
Further, we focused on key agreement and, hence, used \ac{PSK} authentication,
isolating the key exchange effect. It makes our results a lower bound for deployments
whose signatures and certificate chains bear additional load. Both peers also run our
own implementation, barring us from cross-checking the implementation validity against
a different one. At the same time, the shared prerequisites increase the comparability
of different approaches as their implementation affects all equally. Finally,
disruptions to the test environment necessitated restarting configurations. We always
applied the same configuration and manually verified that important parameters like
bandwidth and base latency did not vary across restarts.

Our research compares our own proposals with current alternatives. To reduce that
bias, we decided on metrics and settings early and systematically, before taking any
measurements.

\section{Conclusion}%
\label{sec:conclusion}
The IETF discusses different approaches to integrating \ac{PQC} \acp{KEM} into
IKEv2 but they lack a comparison on one implementation under identical
conditions. We systematically screened the IETF document corpus identifying all
currently discussed approaches (\emph{\ref{rq:ietf_proposals}}), proposed three
novel ones, namely one \ac{FEC}-based, QUIKE, and QUIKE-composite
(\emph{\ref{rq:new_proposals}}), implemented all of them on top of the same
codebase, and evaluated all of them in one test environment emulating various
network conditions and \acp{KEM} (\emph{\ref{rq:performance}}), leading to
$1296$ different configurations and $126\,500$ handshakes. The exact result
data, our implementations, and the IETF screening artifacts are all published
inside our replication package~\cite{repl_package}.

Our results show that the dominant factor for the performance is the
\ac{KEM} size and the transmission approach needs to be adjusted to it. For Classic
McEliece-sized \acp{KEM}, no proposal managed handshakes under
challenging network conditions sufficiently well. For ML-KEM and FrodoKEM, changing
the transport and then using composite \acp{KEM} is more reliable and faster, as well
as leaner to implement, with our proposal QUIKE-composite the fastest. With
datagram-based transport, the existing \ac{ADDKE} RFC~\cite{rfc9370}
requires the largest and most intrusive code change and struggles with more
challenging network environments but can be extended with our \ac{FEC}-mechanism or
IKEv2-level acknowledgments. For those, two concurrent active drafts exist,
with status-ACK~\cite{selective_retransmission} working better than
notify-ACK~\cite{fragment_ack}. We identified gaps for both of them regarding
the necessity to retain superseded IKE \ac{SA} keys and their unspecified retransmission
limit which, like replayable notify-ACKs, opens \ac{DoS} vectors. In the future,
standardization and implementers will need to choose between keeping continuity and
compatibility on the one hand, and adopting simpler, better-performing solutions on
the other hand (\emph{\ref{rq:implications}}).

Future work comprises considering the approaches' impact on authentication. Our tests
used \ac{PSK} authentication. Real-world deployments use public keys and certificate
chains whose migration to \ac{PQC} constitutes its own research problem, also
sparking discussions about implicit \ac{KEM}-based authentication (cf.
Section~\ref{sec:rel_work}) or \ac{EAP}. Investigating those interactions with the
approaches evaluated in this work gives a more holistic view on IKEv2's \ac{PQC}
migration. In terms of standardization, we plan on discussing found gaps of the
acknowledgment drafts in the respective working group. Further, we intend to propose
\ac{FEC} and QUIKE-composite approaches at the IETF\@. During implementation, we
fixed a bug in the QUIC library we build on and intend to upstream this fix. Beyond
that, combining our \ac{FEC} approach with the status-ACK draft and measuring it
against our data would show how well the two optimizations compound. Analyzing how
different loss patterns affect the individual approaches is a further open question.
For QUIKE, implementing the separation of QUIC and IKEv2 into distinct processes and
formally verifying how far that containment limits an attacker able to manipulate the
QUIC implementation are two further steps. Finally, enabling Classic
McEliece-sized keys with acceptable performance and without congestion remains an
important research direction.

\clearpage
\appendix%
\def\theHsection{Appendix.\Alph{section}}%
\section{Measurement Results}%
\label{app:results}
This appendix shows the adjacent data to the figures in Section~\ref{sec:results},
allowing the reader to reproduce the per-configuration distributions.

\subsection{Completion-Time Distributions}%
\begin{figure}[tbp]
	\centering
	\includegraphics[width=\columnwidth]{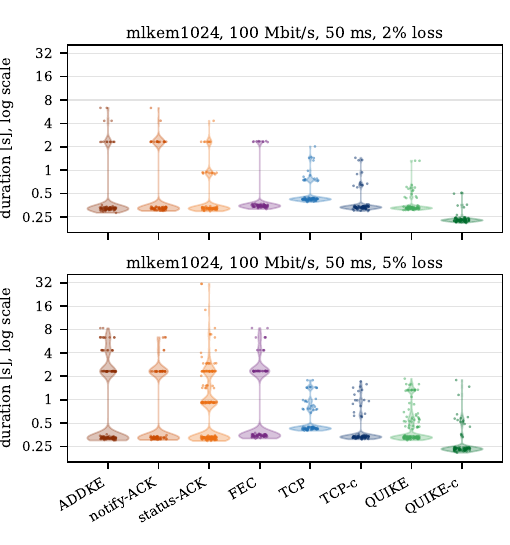}
	\caption{ML-KEM over a $100$\,Mbit/s link at $50$\,ms one-way delay, at
	$2\%$ and $5\%$ packet loss.}%
	\label{fig:app_violin_mlkem}
\end{figure}

\begin{figure}[ptb]
	\centering
	\includegraphics[width=\columnwidth]{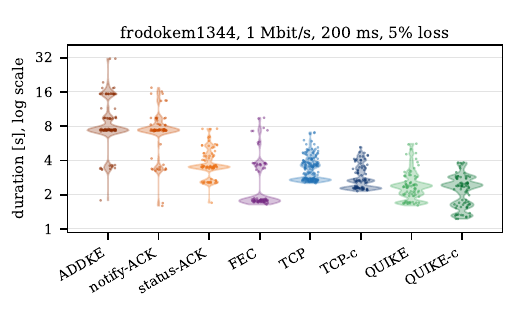}
	\caption{FrodoKEM over a $1$\,Mbit/s link at $200$\,ms one-way delay and
	$5\%$ packet loss.}%
	\label{fig:app_violin_frodo}
\end{figure}

\begin{figure}[ptb]
	\centering
	\includegraphics[width=\columnwidth]{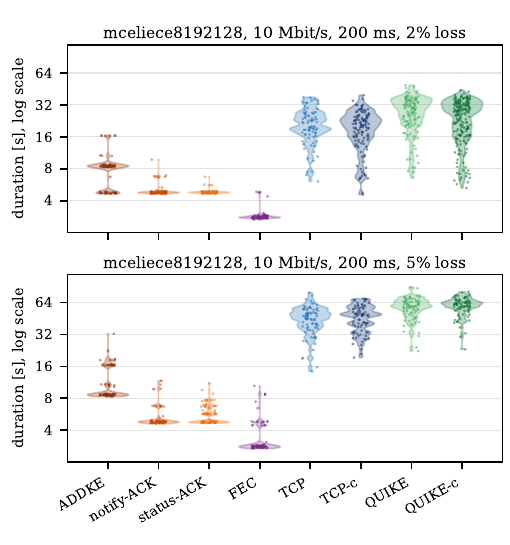}
	\caption{Classic McEliece over a $10$\,Mbit/s link at $200$\,ms one-way
	delay, at $2\%$ and $5\%$ packet loss.}%
	\label{fig:app_violin_mceliece}
\end{figure}

\label{app:violin}
Figure~\ref{fig:res_distributions} shows one FrodoKEM configuration. Figures~\ref{fig:app_violin_mlkem}--\ref{fig:app_violin_mceliece}
extend it to the other \acp{KEM}, further loss levels, and different network speeds,
showing the deductions' general applicability.

\subsection{Transmission Overhead of the Smaller KEMs}%
\begin{figure}[t]
	\centering
	\includegraphics[width=\columnwidth]{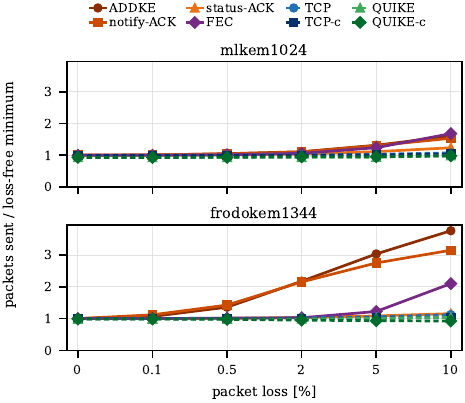}
	\caption{Packets the initiator sends over a $10$\,Mbit/s link for ML-KEM and
		FrodoKEM, each approach relative to itself on an uncongested loss-free path (cf.
		Figure~\ref{fig:res_overhead}).}%
	\label{fig:app_overhead_small}
\end{figure}

\label{app:efficiency}
Figure~\ref{fig:res_overhead} shows the efficiency of Classic McEliece, where the approaches'
overhead differs most. ML-KEM and FrodoKEM reproduce the same ordering
but at smaller magnitudes as shown in Figure~\ref{fig:app_overhead_small}.

\subsection{The Complete Grid}%
\begin{figure*}[!p]
	\centering
	\includegraphics[width=\textwidth]{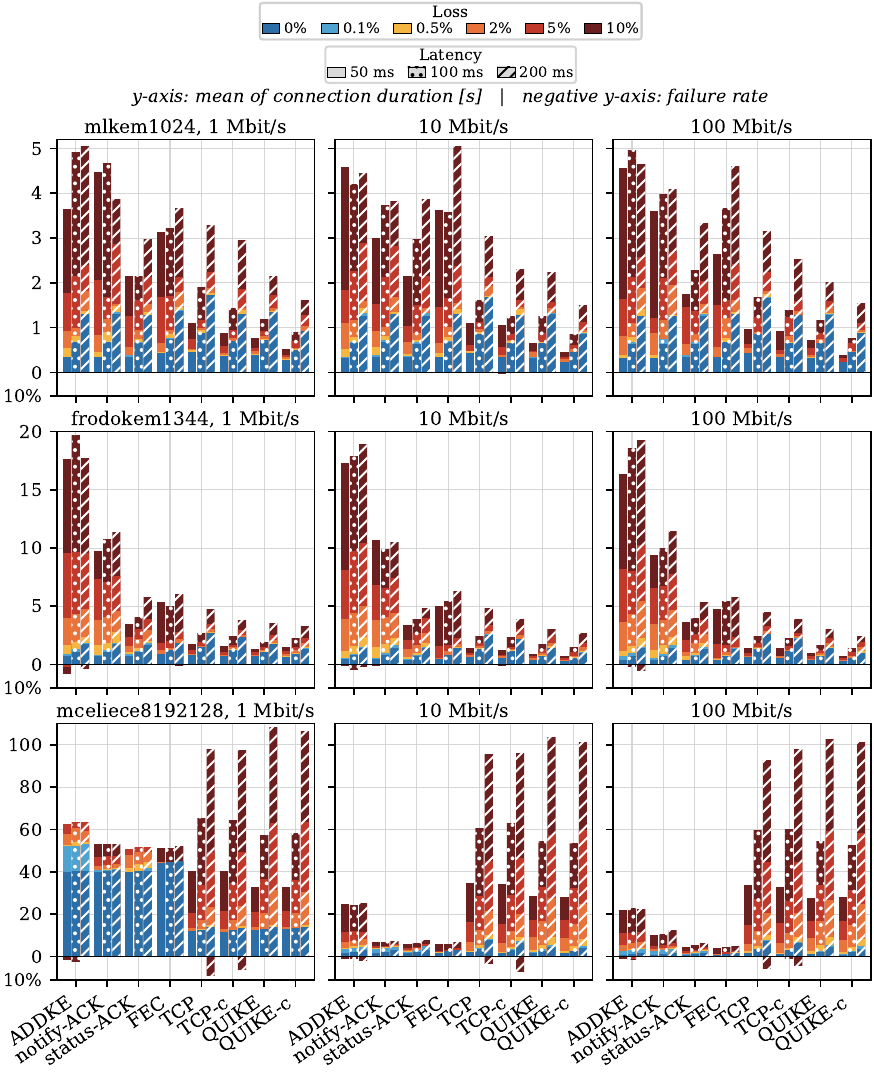}
	\caption{Mean handshake duration over the complete grid. Rows are the
	\acp{KEM}, each with its own vertical axis shared across the row, and
	columns are the bandwidths.}%
	\label{fig:app_grid}
\end{figure*}

\label{app:stacked}
Figure~\ref{fig:app_grid} gives every measured configuration. Per approach
there is one bar per one-way delay, loss levels behind one another, each showing its
own excess over the next lower loss level. The mirrored area below zero depicts the
failure rate.

\clearpage

\section*{Acknowledgment}
The authors used Anthropic's Claude with the models Fable5, Opus4.8, Opus5, and Opus5.5 during
the preparation of this article. It assisted in writing the \acs{PoC} implementation,
revising the manuscript for internal consistency and grammar, and writing the scripts
for the analysis, systematic search, and plotting of the result data. The authors
verified all analyses against the recorded measurements, reviewed and edited every
generated passage, and take full responsibility for the content of this publication.

\bibliographystyle{IEEEtran}
\bibliography{bibliography}

\unless\ifanonymous
  \begin{IEEEbiography}[{\includegraphics[width=1in,height=1.25in,clip,keepaspectratio]{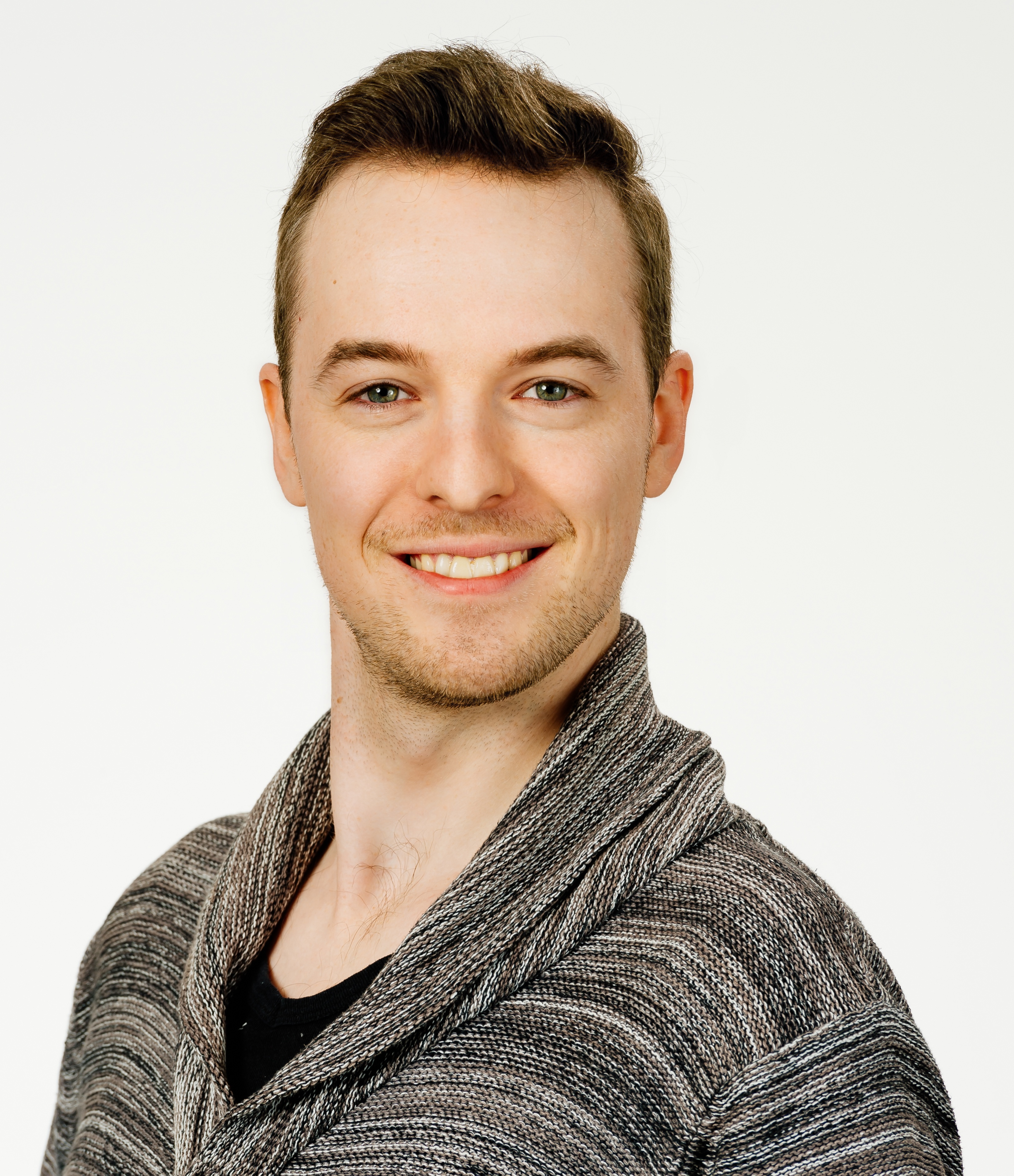}}]
{Daniel Herzinger} holds a Master of Science in Computer Science.
He started working on post-quantum cryptography (PQC) and its adoption into real-world scenarios during his Master's thesis in 2020, together with the German IT security company genua, where he started his IT security journey in 2015. After finishing his studies, he first dove into supporting customers with securing their IT and OT infrastructures at genua's consulting department for two years. He then indulged in his passion for PQC by joining the research group in 2023, focusing on the transition of complex systems and infrastructures to quantum resistance.
\end{IEEEbiography}

\begin{IEEEbiography}[{\includegraphics[width=1in,height=1.25in,clip,keepaspectratio]{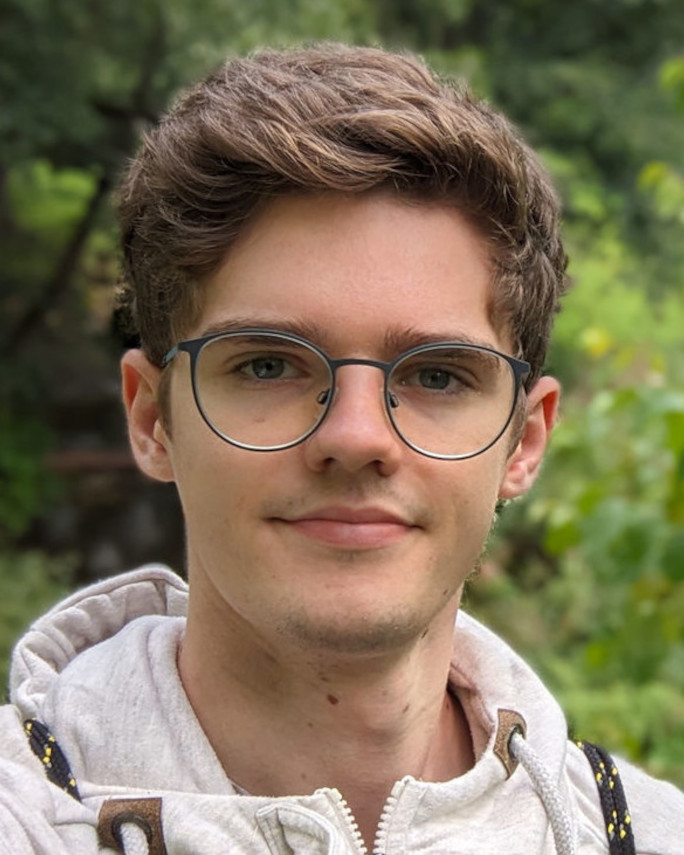}}]
{Marco Thomas} holds a Master of Science in IT Security.
He joined the German IT security company genua GmbH in 2019 as a student and finished his studies in 2026 with his Master's thesis about real-world usability of post-quantum cryptography (PQC) in virtual private network (VPN) software.
Throughout his studies, he focused on understanding deeply technical computer science topics, such as networking and kernel development. After finishing his studies, he joined the development department of genua, which is in charge of developing VPN technologies using
 the IKEv2 protocol. There, he works on integrating post-quantum cryptography into these products.
\end{IEEEbiography}

\begin{IEEEbiography}[{\includegraphics[width=1in,height=1.25in,clip,keepaspectratio]{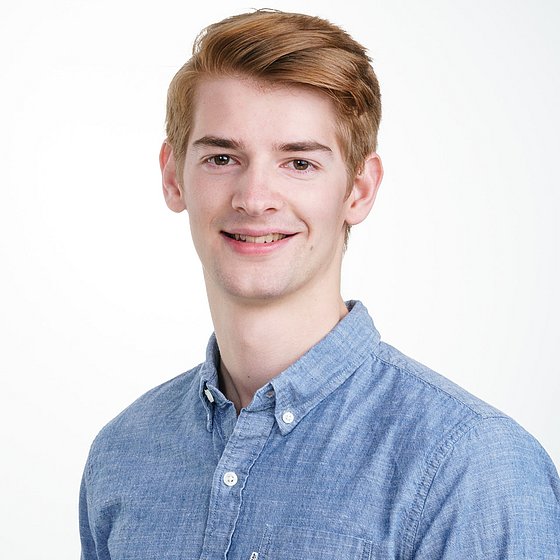}}]
	{Tobias Handirk} is a researcher for Post-Quantum Cryptography at the German IT security company genua GmbH. He joined the company in 2025, after completing a PhD in Computer Science with \textit{summa cum laude}, which he obtained in 2024 at the Bergische Universität Wuppertal. In his PhD research he worked on the provable security of real-world cryptographic protocols, with a particular focus on key exchange protocols. Prior to that, he finished a Master of Science in Computer Science at the University of Paderborn with a specialization in the area ``Networks and Communications''.
\end{IEEEbiography}

\begin{IEEEbiography}[{\includegraphics[width=1in,height=1.25in,clip,keepaspectratio]{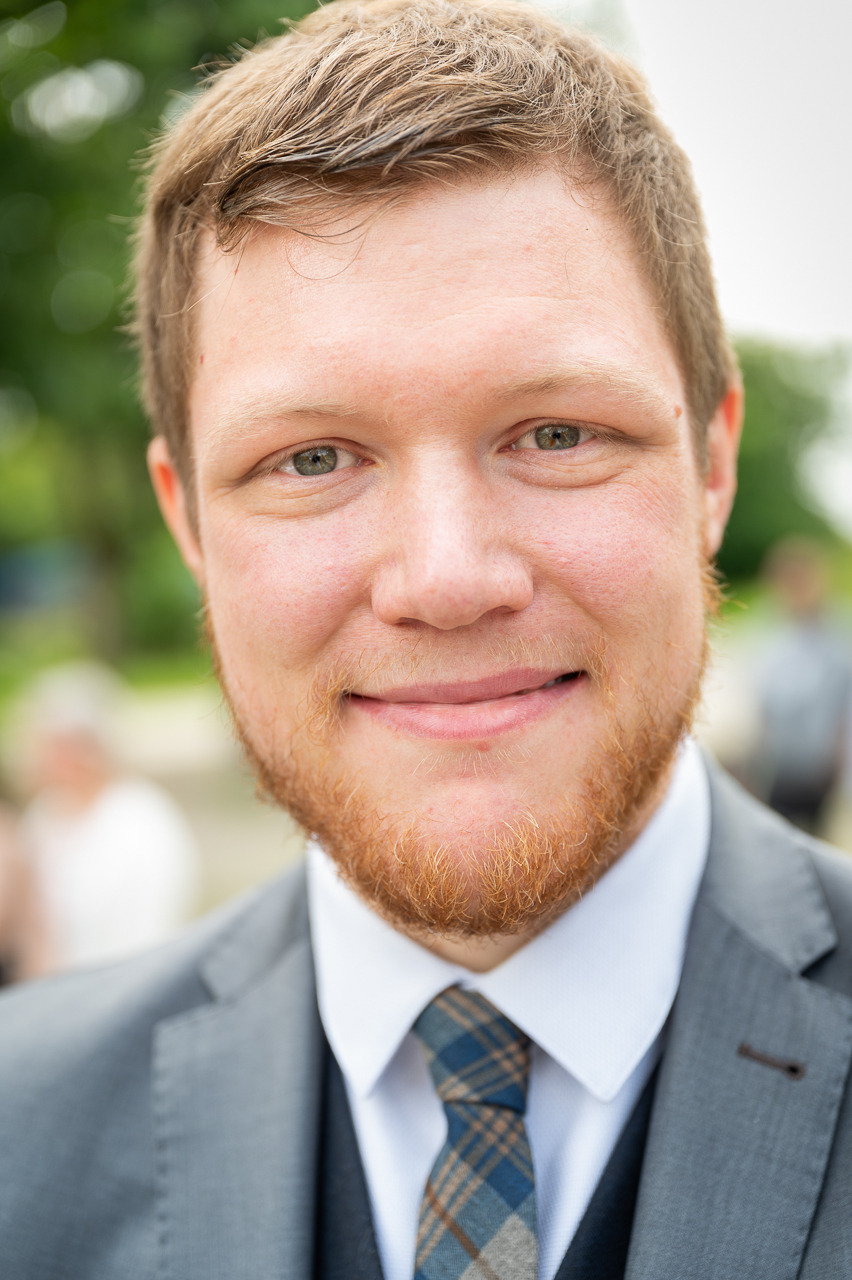}}]
  {Stefan-Lukas Gazdag} holds a Master of Science in Computer Science. He has focused on post-quantum cryptography (PQC) since 2013, overcoming obstacles in practical use. At the German IT security company genua’s research group he works with his team on publicly funded PQC research projects and other collaborations, enabling the transition to quantum-safe networks. He was, e.g., involved in the publication of the first explicit PQC RFC, RFC 8391, describing the signature scheme XMSS, recommended, e.g., by US NIST, NSA and German BSI.
  \end{IEEEbiography}

\begin{IEEEbiography}[{\includegraphics[width=1in,height=1.25in,clip,keepaspectratio]{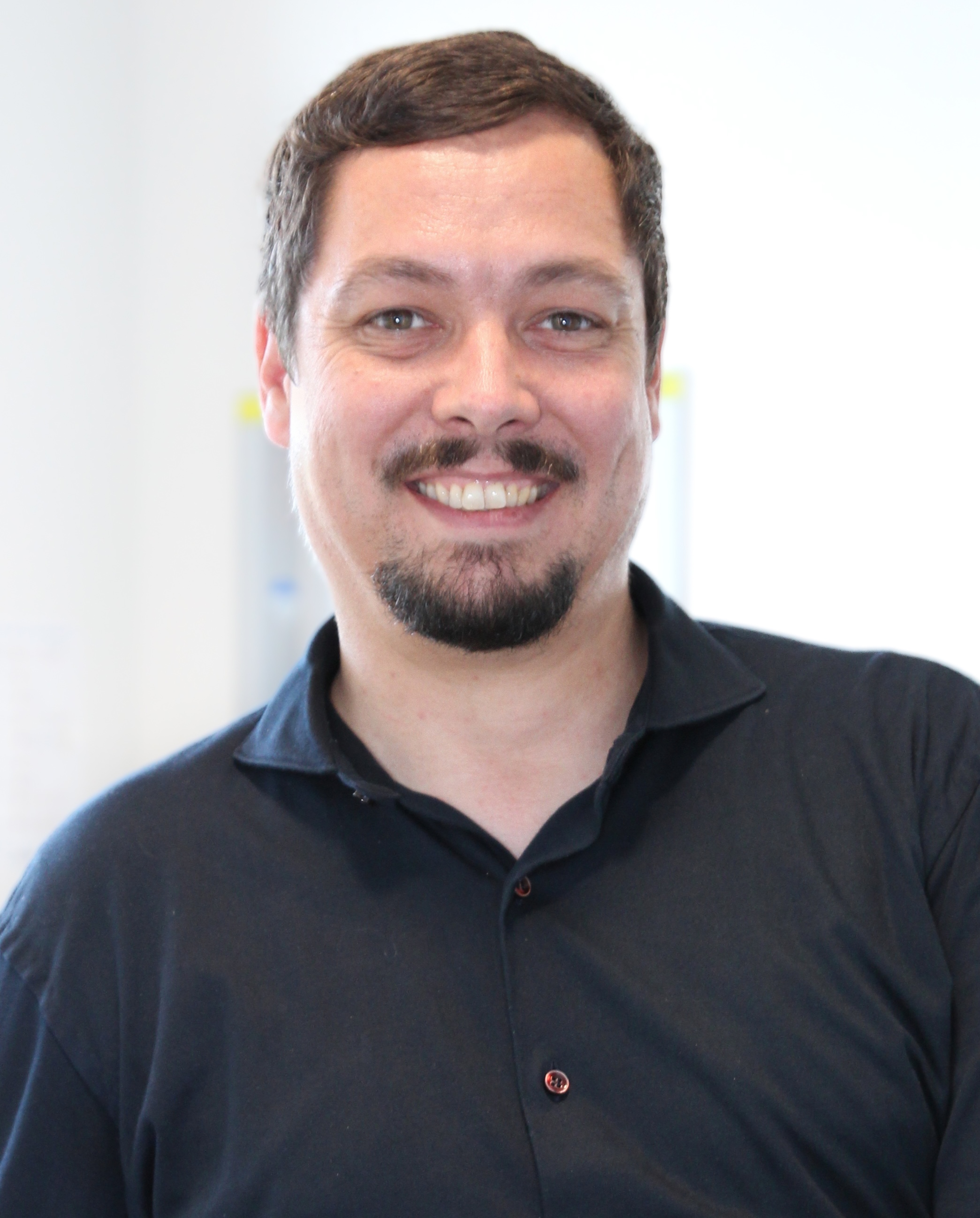}}]
{Daniel Loebenberger} received his doctorate in cryptography from the University of Bonn in 2012, where he conducted research and taught until the end of 2015. From 2016 to 2019, he worked as an IT security expert with a focus on cryptography at genua GmbH, a subsidiary of the Bundesdruckerei GmbH, the German Federal Printing Office, on various topics of professional high-security components. Daniel Loebenberger was appointed Professor of Cybersecurity at OTH Amberg-Weiden University of Applied Sciences in January 2019 and has also headed the Secure Infrastructure department at Fraunhofer AISEC's Weiden site since then. His group focuses on applied post-quantum cryptography and quantum-safe infrastructures. Since 2026, he has also been an Honorary Visiting Professor at Nanyang Technological University, Singapore.
\end{IEEEbiography}

\begin{IEEEbiography}[{\includegraphics[width=1in,height=1.25in,clip,keepaspectratio]{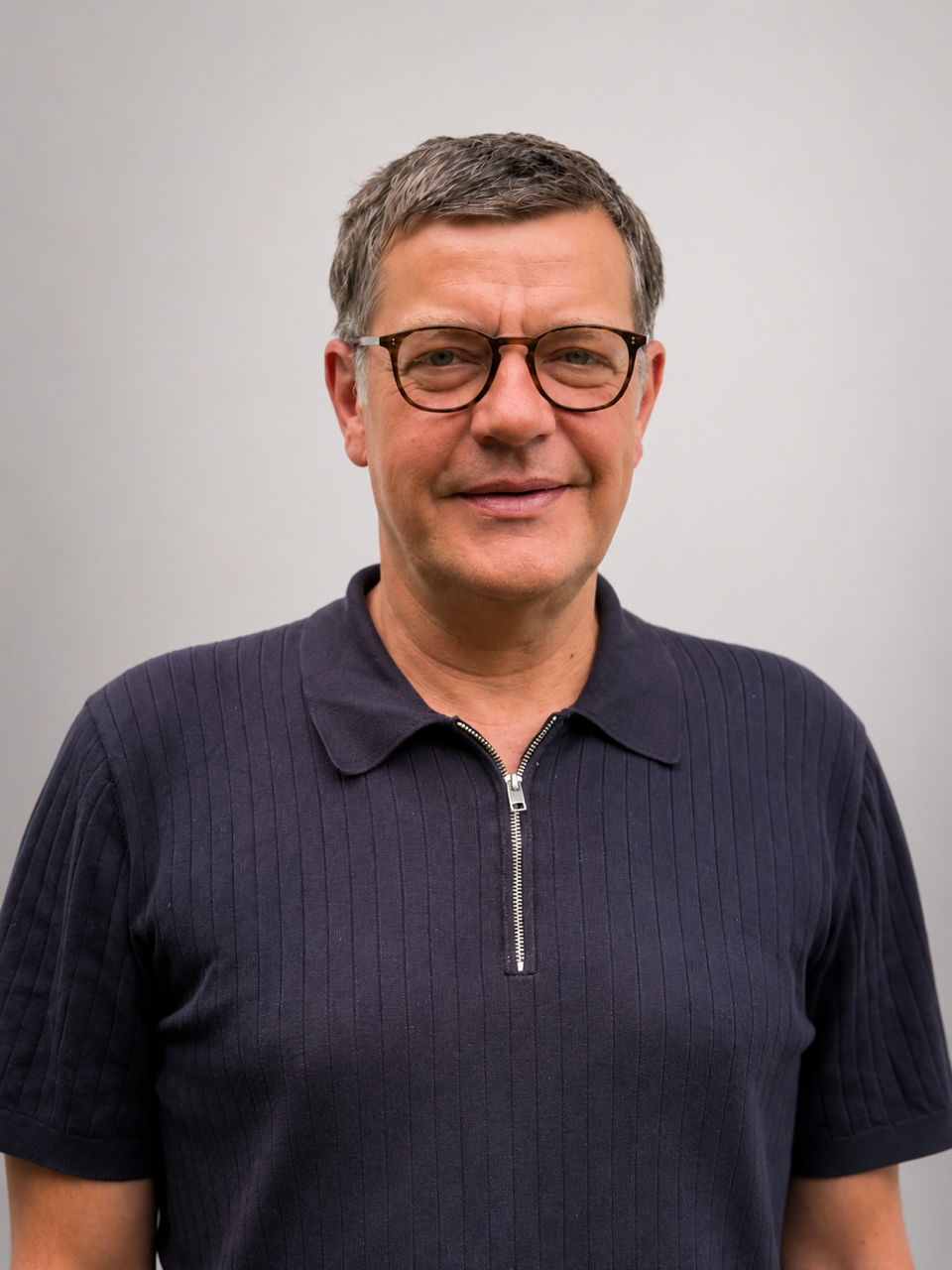}}]
	{Marian Margraf} is a professor of information security at Freie Universität Berlin
	and head of department at the Fraunhofer Institute AISEC\@. His research group
	focuses on usable privacy and security, cryptography---particularly post-quantum
	cryptography---and electronic identities. He has more than 20 years of experience in the field of information security. He began his career from 2003 to 2008 as a cryptologist at the Federal Office for Information Security, where he helped develop the cryptographic protocols for electronic passports and ID cards. In 2008, he moved to the Federal Ministry of the Interior, where he developed information security strategies.
\end{IEEEbiography}

\fi

\EOD

\end{document}